\documentclass{amsart}
\usepackage{graphicx} 
\usepackage{hyperref}
\usepackage{nameref,hyperref}
\usepackage{xcolor}
\hypersetup{allbordercolors=lightgray}
\usepackage{url}
\usepackage{subfig}
\usepackage{morefloats}
\usepackage{multirow}
\usepackage{palatino}
\usepackage{eucal}
\usepackage{todonotes}
\usepackage{setspace}
\usepackage{cite}
\usepackage{pbox}

\makeatletter
\renewcommand{\@biblabel}[1]{\quad#1.}
\makeatother
\pdfoutput=1

\newcommand{\imgt}{IMGT}
\newcommand{\tigger}{TIgGER}
\newcommand{\igdiscover}{IgDiscover}

\newcommand{\nsnp}{\ensuremath{N_{\text{snp}}}}
\newcommand{\nth}{$^{\text{th}}$}
\newcommand{\simGlsTrees}{~\ref{FIGglsSimFull},~\ref{FIGglsSimIgdiscover},~\ref{FIGglsSimTigger},~\ref{FIGglsSimPartis}}
\newcommand{\methodVsMethodFigures}{~\ref{FIGdataJasonMGHDmethodVsMethod},~\ref{FIGdataJasonInfluenzaMethodVsMethod},~\ref{FIGdataShengGSSPMethodVsMethod},~\ref{FIGdataJasonMGARmethodVsMethod},~\ref{FIGdataJasonMGMKmethodVsMethod}}
\newcommand{\sampleVsSampleFigures}{~\ref{FIGdataJasonInfluenzaSampleVsSample},~\ref{FIGdataShengGSSPsampleVsSample}}
\newcommand{\zenodolink}{\url{https://zenodo.org/record/1233698}}

\newcommand{\forarxiv}[1]{#1}
\newcommand{\notforarxiv}[1]{}

\newcommand{\beginsupplement}{%
        \setcounter{table}{0}
        \renewcommand{\thetable}{S\arabic{table}}%
        \setcounter{figure}{0}
        \renewcommand{\thefigure}{S\arabic{figure}}%
     }

\begin{document}

\vspace*{0.35in}

\begin{flushleft}
{\Large
\textbf\newline{Per-sample immunoglobulin germline inference from B cell receptor deep sequencing data}
}
\newline

Duncan K. Ralph\textsuperscript{1,*},
Frederick A. Matsen IV\textsuperscript{1}

\bigskip
\bf{1} Fred Hutchinson Cancer Research Center, Seattle, Washington, USA
\bigskip

\textsuperscript{*} dkralph@gmail.com

\end{flushleft}

\section*{Abstract}
The collection of immunoglobulin genes in an individual's germline, which gives rise to B cell receptors via recombination, is known to vary significantly across individuals.
In humans, for example, each individual has only a fraction of the several hundred known V alleles.
Furthermore, the currently-accepted set of known V alleles is both incomplete (particularly for non-European samples), and contains a significant number of spurious alleles.
The resulting uncertainty as to which immunoglobulin alleles are present in any given sample results in inaccurate B cell receptor sequence annotations, and in particular inaccurate inferred naive ancestors.
In this paper we first show that the currently widespread practice of aligning each sequence to its closest match in the full set of \imgt\ alleles results in a very large number of spurious alleles that are not in the sample's true set of germline V alleles.
We then describe a new method for inferring each individual's germline gene set from deep sequencing data, and show that it improves upon existing methods by making a detailed comparison on a variety of simulated and real data samples.
This new method has been integrated into the partis annotation and clonal family inference package, available at \url{https://github.com/psathyrella/partis}, and is run by default without affecting overall run time.

\section*{Author Summary}
Antibodies are an important component of the adaptive immune system, which itself determines our response to both pathogens and vaccines.
They are produced by B cells through somatic recombination of germline DNA, which results in a vast diversity of antigen binding affinities across the B cell repertoire.
We typically learn about the development of this repertoire, and its history of interaction with antigens, by sequencing large numbers of the DNA sequences from which antibodies are derived.
In order to understand such data, it is necessary to determine the combination of germline V, D, and J genes that was rearranged to form each such B cell receptor sequence.
This is difficult, however, because the immunoglobulin locus exhibits an extraordinary level of diversity across individuals -- encompassing both allelic variation and gene duplication, deletion, and conversion -- and because the locus's large size and repetitive structure make germline sequencing very difficult.
In this paper we describe a new computational method that avoids this difficulty by inferring each individual's set of immunoglobulin germline genes directly from expressed B cell receptor sequence data.

\section*{Introduction}
The heavy and light chain B cell receptor (BCR) loci arise from a random recombination of germline V, D, and J genes.
Repeated across many B cells, this generates the vast diversity of naive BCRs that is integral to the adaptive immune system.
As an additional source of population-wide variation, there is significant variation of germline genes between individuals.
Databases such as \imgt~\cite{Lefranc2009-iu} aim to collect and organize this ensemble of germline genes.

The analysis of BCR sequence data begins with the alignment of each sequence against a set of germline V, D, and J genes.
A variety of methods (e.g.\ \cite{Ye2013-ei,imgt-high-v-quest,partis-annotation,Bolotin2015-zb}) have been developed to accomplish the basic task of deciding which V, D, and J genes gave rise to each observed sequence.
There has been less work, however, toward measuring the extent to which the set of germline genes used for this analysis resembles the germline gene set actually present in the individual from which the sequence data was derived.
Most methods simply use the full set of germline genes from a database such as \imgt~\cite{Lefranc2009-iu} for all samples.

One problem with this approach is that the \imgt\ set includes genes from all individuals of a species, while any single individual's germline contains only a fraction of these (roughly 50 out of 250 V genes, 25 of 35 D, and 6 of 12 J).
This is problematic for sequencing studies that use antigen-experienced B cells that have been through several rounds of somatic hypermutation (SHM), which obscures the identity of the original germline gene.
As we show below, this leads to large numbers of spurious gene assignments, and an inferred germline gene set with many more alleles than are in the individual's true set.

Another problem with this approach is that no database contains a perfect catalog of the complete immunoglobulin germline diversity of each species.
Sequencing continues to uncover novel human V genes that are not in any previous database~\cite{Wang2011-ns,boyd2010-yj,kidd2012-cj,Watson2013-heavy,Watson2014-light,tigger,igdiscover,Scheepers2015-no}.
Additionally, a significant fraction of the sequences in existing databases are likely the result of sequencing error rather than real biological variation~\cite{Lee2006-ti,Lee2007-ej,Wang2008-gp}.
Our knowledge of the immunoglobulin locus is even less complete for other species~\cite{Collins2015-gm,igdiscover}.

Improving our understanding of the immunoglobulin locus, however, is not simply a matter of applying standard genome sequencing protocols more broadly.
Most genome sequencing is performed on lymphoblastoid cell lines~\cite{Venter2001-vo,pmid12466850,1000_Genomes_Project_Consortium2010-rc}, whose prior rearrangement has destroyed much of the information about the original immunoglobulin locus.
The obvious solution would be to sequence other cell types; however assembly challenges due to the complexity and repetitiveness of the locus~\cite{Treangen2011-ch} mean that even sequencing an intact immunoglobulin locus is not straightforward.
The IGHV locus, for instance, consists of about 120 V genes, roughly two-thirds of which are non-functional pseudogenes, spread over a megabase of chromosome 14~\cite{Watson2013-heavy}.
The immunoglobulin locus is also subject to widespread gene duplication, deletion, and conversion~\cite{boyd2010-yj,kidd2012-cj,Chimge2005-ws,Luo2017-pu}.
Thus although databases such as the 1000 Genomes project and the Simons Genome Diversity Project can be used to investigate immunoglobulin diversity~\cite{Yu2017-aq,Luo2017-pu}, this approach is not without pitfalls~\cite{Watson2017-fg}.

Discrepancies between a BCR-sequenced individual's true set of germline genes and the set used to analyze their BCR sequences cause a number of practical problems.
First, finding associations between particular germline genes and an immunological response is difficult if the gene assignment itself is suspect.
This would impact, for example, recent work on the effects of the presence or absence of individual alleles on broadly neutralizing anti-influenza antibody development~\cite{Avnir2016-ax}.
Second, such misassignment leads to inaccurate inferred naive ancestor sequences.
Efforts to synthesize these inaccurate ancestral sequences in the lab and study their binding properties may then result in erroneous conclusions, since even single amino acid changes can have large effects on affinity~\cite{Liu2003-kg}.
And finally, studies of mutation~\cite{Rogozin1992-xv,Cui2016-rs} and selection~\cite{Yaari2012-kk,McCoy2015-qi} during affinity maturation depend upon accurate inferred naive sequences in order to correctly identify somatic mutations.

Our current understanding of the immunoglobulin locus comes largely from a small number of low-throughput genome and BAC library sequencing studies.
The first complete sequence of the locus~\cite{Matsuda1998-tw}, which has been included in the first few drafts of the human genome, was assembled from several different cell lines and is therefore not a haplotype.
More recently, a single complete haplotype of the heavy~\cite{Watson2013-heavy} and light~\cite{Watson2014-light} chain loci has been published.
In addition to these larger efforts, many less-comprehensive studies of the locus have been cataloged at \url{www.imgt.org}.

Advances in sequencing technology, however, have allowed progress to come also from inference on expressed BCR repertoires.
Several initial studies inferred germline sets by combining computational analysis with expert scrutiny, with one paper reporting a high level of diversity with many novel (non-\imgt) alleles across 12 individuals~\cite{boyd2010-yj}, and a second extending those results to 18 complete haplotypes~\cite{kidd2012-cj}.
Similar work by a different team used naive sequences to infer germline sets and haplotype linkage information for two individuals~\cite{Elhanati2015-ld}.
None of these studies, however, resulted in a generally-applicable software package or included a broad-scale validation of their methods.

More recently, software packages have been developed that enable fully-automated germline inference including novel allele discovery.
\tigger~\cite{tigger} uses a detailed per-position fitting procedure to find new alleles separated by a small number of point mutations from genes in a known database, and a heuristic prevalence threshold-based procedure to infer germline sets.
The \igdiscover\ package~\cite{igdiscover} infers germline sets using Levenshtein distance-based hierarchical UPGMA clustering on low-SHM IgM samples.
This approach allows \igdiscover\ to find new alleles separated by an arbitrary number of point mutations and insertion/deletion events, and frees it from the need for an initial species-specific starting database.

In this paper we present a new method for automated inference of per-sample germline V gene sets from expressed BCR sequence data.
We first compare our method's accuracy on a variety of simulated samples both to the common practice of aligning against the full \imgt\ set, and to the two existing germline inference methods, \tigger\ and \igdiscover.
We find that use of the full \imgt\ set results in a very large number of spuriously-inferred alleles on typical samples, as well as inaccurately inferred naive sequences.
We further find that while our method infers a similar fraction of correct and incorrect genes as \tigger\ and \igdiscover, its inferred genes are more similar to the true genes, and thus our method's inferred naive sequences are significantly more accurate.
We then use a variety of real data samples from the literature to compare the germline gene sets inferred by our method to those from \tigger\ and \igdiscover, and find no reason to believe that our simulation does not accurately mimic real data.
Because our method performs well on samples with elevated levels of SHM, it is more generally applicable than \igdiscover, which is restricted to low-SHM IgM samples.
In addition, our method is run by default in the general-purpose partis package, which also provides annotation, clonal family inference, and simulation, while \tigger\ and \igdiscover\ must be run as separate steps.
Because the D and J loci vary much less between individuals than does V (and because D inference would be very challenging) in this paper we follow these other software packages in limiting ourselves to studies of V diversity.

Because of the high prevalence of both single nucleotide polymorphisms (SNPs) and structural variants in the immunoglobulin locus, there is no single reference genome to which all variants can be mapped, and thus standard SNP nomenclature appears insufficient.
In this paper the usage of ``gene'' and ``allele'' is thus largely interchangeable.
In addition, we define the ``germline haplotype'' as the set of germline genes on a single chromosome, while ``germline gene set'' refers to the full set on both the maternal and paternal chromosomes.
In cases where confusion is unlikely, the latter will be shortened to ``germline set''.

\section*{Results}
\newcommand{\varvalfigs}{Figs~\ref{FIGvarval1} and~\ref{FIGvarval2}}

\subsection*{Simulation methods summary}
In order to establish an expectation for how germline inference methods will perform on real data, we first investigate performance on a number of simulation samples.
BCR repertoires differ significantly in many different variables such as SHM levels, germline set complexity, and clonal family structure.
Although we would in principle like to explore germline inference accuracy by varying all of these variables simultaneously, this is combinatorially infeasible, and we thus adopt a two-stage approach to validation.
We first vary one variable at a time, while holding all others constant, using simplified ``sparse'' repertoires consisting of sequences stemming from only a few genes.
We then choose several representative values for each variable, and simulate full, realistic repertoires at these values.
Geometrically, this can be imagined as investigating performance first along many slices through the parameter space, and then at several fixed points.
This approach is motivated by the fact that, in sequence-similarity space, realistic repertoires are composed of widely-spaced groups of genes, where each group consists of a few genes that are much closer to each other than the typical between-group spacing.
The genes within each group are thus easily confused with each other due to SHM, but not with genes in other groups.
The sparse repertoires effectively recreate the dynamics within such a group, while allowing exploration of a much larger portion of parameter space than if we were to use full repertoires for all simulations.

In these simulations, the germline set for each sparse repertoire consists of one known germline gene, and either one or two novel alleles.
Each full-repertoire sample, meanwhile, is generated by choosing a number of V, D, and J genes, and some number of alleles for each of these genes, based on results from germline sequencing studies (see Methods), which results in roughly 55 V, 25 D, and 6 J alleles per sample.

\subsection*{Validation results}
\newcommand{\varvalextratext}{
  {\bf Fraction of true alleles missing (left) and alleles spuriously inferred (right)} by partis on simplified ``sparse'' repertoires as a function of the number of sequences in the sample.
  Each point represents the mean performance ($\pm$ standard error) on 50 independent simulation samples of the indicated sample size varying the following variables.
}

\begin{figure}[!ht]
\forarxiv{\includegraphics[width=5in]{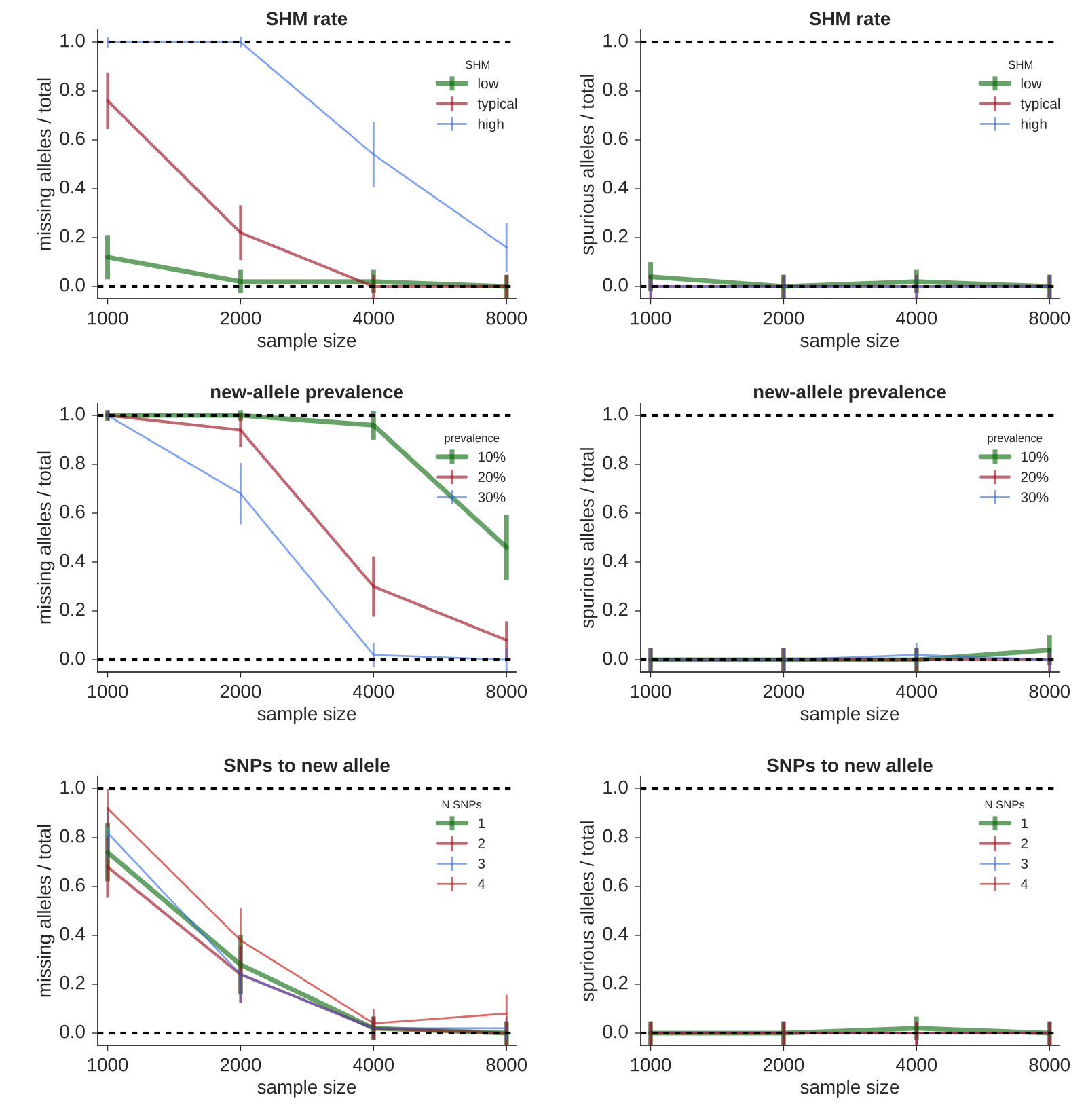}}
\caption{\
  \varvalextratext
  {\bf Top:} SHM levels (the SHM distributions corresponding to ``low'', ``typical'', and ``high'' are shown in Fig~\ref{FIGvFreqDistr}).
  {\bf Middle:} new-allele prevalence (as a fraction of the existing allele's prevalence). 
  {\bf Bottom:} number of SNPs (\nsnp) separating new and existing alleles.
}
\label{FIGvarval1}
\end{figure}

\begin{figure}[!ht]
\forarxiv{\includegraphics[width=5in]{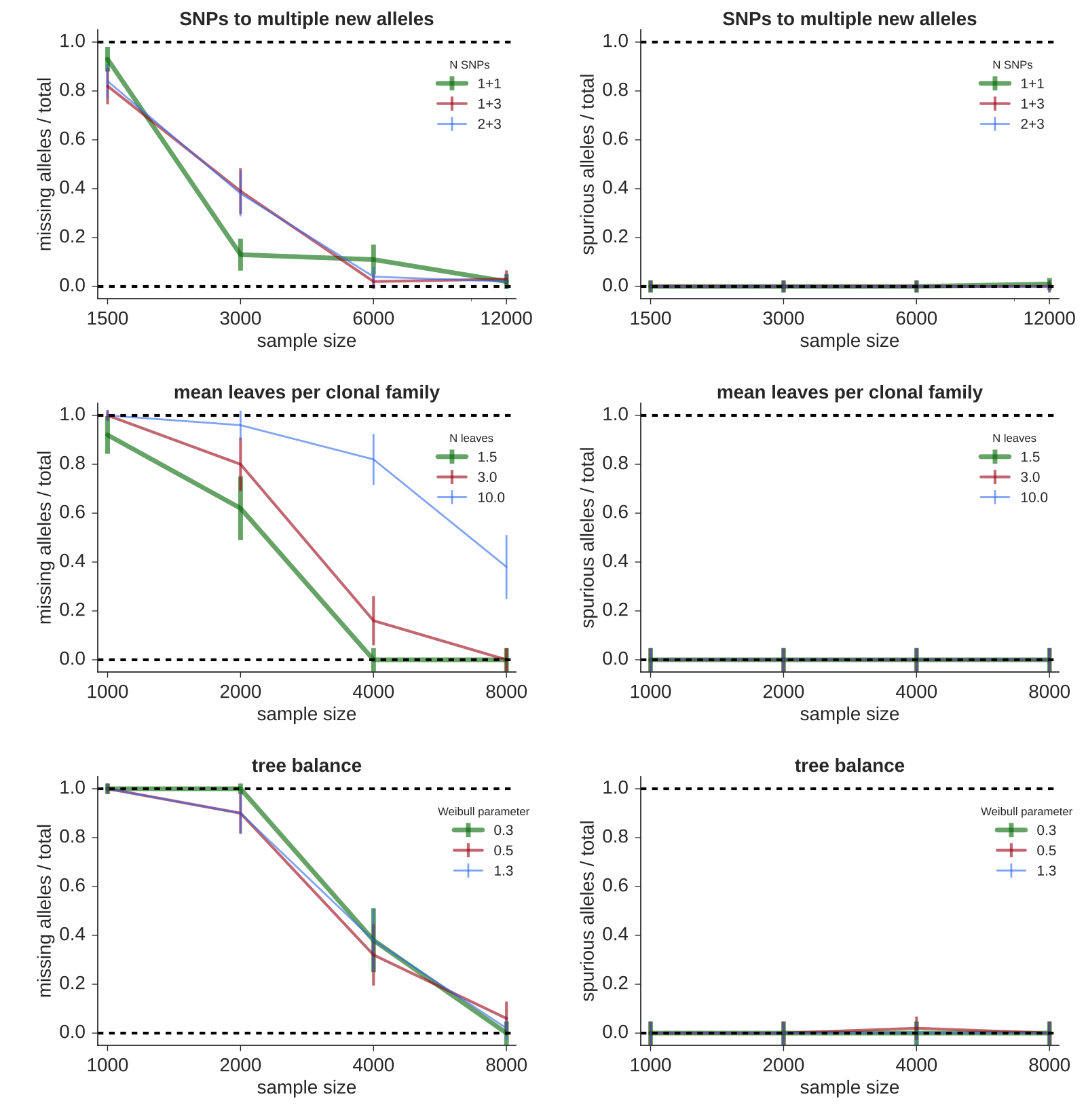}}
\caption{\
  \varvalextratext
  {\bf Top:} \nsnp\ with multiple new alleles, where, e.g.  ``1 + 3'' indicates two new alleles, separated by 1 and 3 SNPs from the same existing allele.  
  {\bf Middle:} mean number of leaves per clonal family.
  {\bf Bottom:} tree balance. 
}
\label{FIGvarval2}
\end{figure}


\newcommand{\glssamples}{Shown on the first three replicates (0-2) of both the low-SHM (left), and high-SHM (right) full-repertoire simulation samples (see text).}
\newcommand{\blBlurb}{Branch lengths connecting different V gene families are set to zero.}  

\begin{figure}[!ht]
\forarxiv{\includegraphics[width=5.5in]{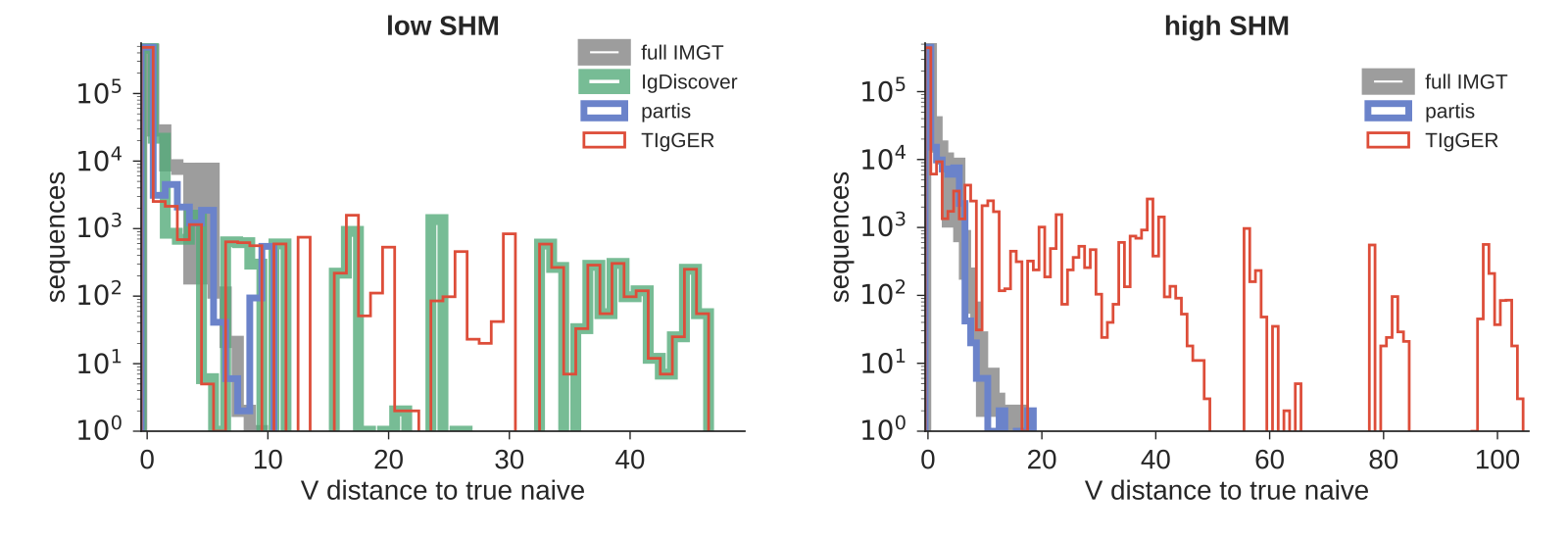}}

{\small
  \centering
\begin{tabular}{lcc}
& low SHM & high SHM\\
\hline
full \imgt   &  $0.22 \pm 0.02$   &  $0.31 \pm 0.02$  \\
\igdiscover  &  $0.36 \pm 0.08$   & \\
partis       &  $0.08 \pm 0.02$   &  $0.27 \pm 0.03$  \\
\tigger      &  $0.42 \pm 0.07$   &  $1.78 \pm 0.36$  \\
\hline
\end{tabular}}

\caption{\
  {\bf Full-repertoire V naive accuracy (Hamming distance between true and inferred V naive sequences) for the three germline inference methods plus ``full \imgt'' annotation.}
  Results are the sum (figures, top) or mean (table, $\pm$ standard error) of ten independent 50,000-sequence samples for both low-SHM (left) and high-SHM (right).
  \igdiscover\ is shown for only the low-SHM samples, since it is designed only for IgM.
}
\label{FIGglsSimAnnotationPerformance}
\end{figure}

\begin{table}[!ht]
{\small
  \centering
\begin{tabular}{lccc|ccc}
&  & low SHM   &  &  & high SHM & \\
&  \# missing  &  \# spurious  &  \# correct  &  \# missing  &  \# spurious  &  \# correct \\
\hline
full \imgt   &  $ 5.0 \pm 0.0 $   &   $53.0 \pm 2.4 $   &   $50.3 \pm 0.7 $   &  $ 5.0 \pm 0.0 $   &   $80.3 \pm 3.5 $   &   $51.5 \pm 1.0 $  \\
\igdiscover  &  $ 3.7 \pm 0.2 $   &   $ 3.4 \pm 0.3 $   &   $51.6 \pm 0.6 $   & & & \\
\tigger      &  $ 2.7 \pm 0.3 $   &   $ 0.4 \pm 0.2 $   &   $52.6 \pm 0.6 $   &  $ 7.8 \pm 0.4 $   &   $ 0.0 \pm 0.0 $   &   $48.7 \pm 0.9 $  \\
partis       &  $ 2.4 \pm 0.4 $   &   $ 1.3 \pm 0.3 $   &   $52.9 \pm 0.9 $   &  $ 9.3 \pm 0.7 $   &   $ 2.9 \pm 0.3 $   &   $47.2 \pm 0.7 $  \\
\hline
\end{tabular}
\caption{\
  {\bf Missing and spurious alleles on full-repertoire simulation for the three germline inference methods plus ``full \imgt'' annotation}.
  Results are the mean ($\pm$ standard error) of ten independent 50,000-sequence samples for both low-SHM (left) and high-SHM (right).
  The columns are {\bf \# missing:} mean number of true alleles missing from the inferred germline set, {\bf \# spurious:} the number inferred that are not in the true germline set, and {\bf \# correct:} the number in common between the inferred and true germline sets.
  \igdiscover\ is shown for only the low-SHM samples, since it is designed only for IgM.
}\label{TABLEglsSimSummary}}
\end{table}

\begin{figure}[!ht]
\forarxiv{\includegraphics[width=5in]{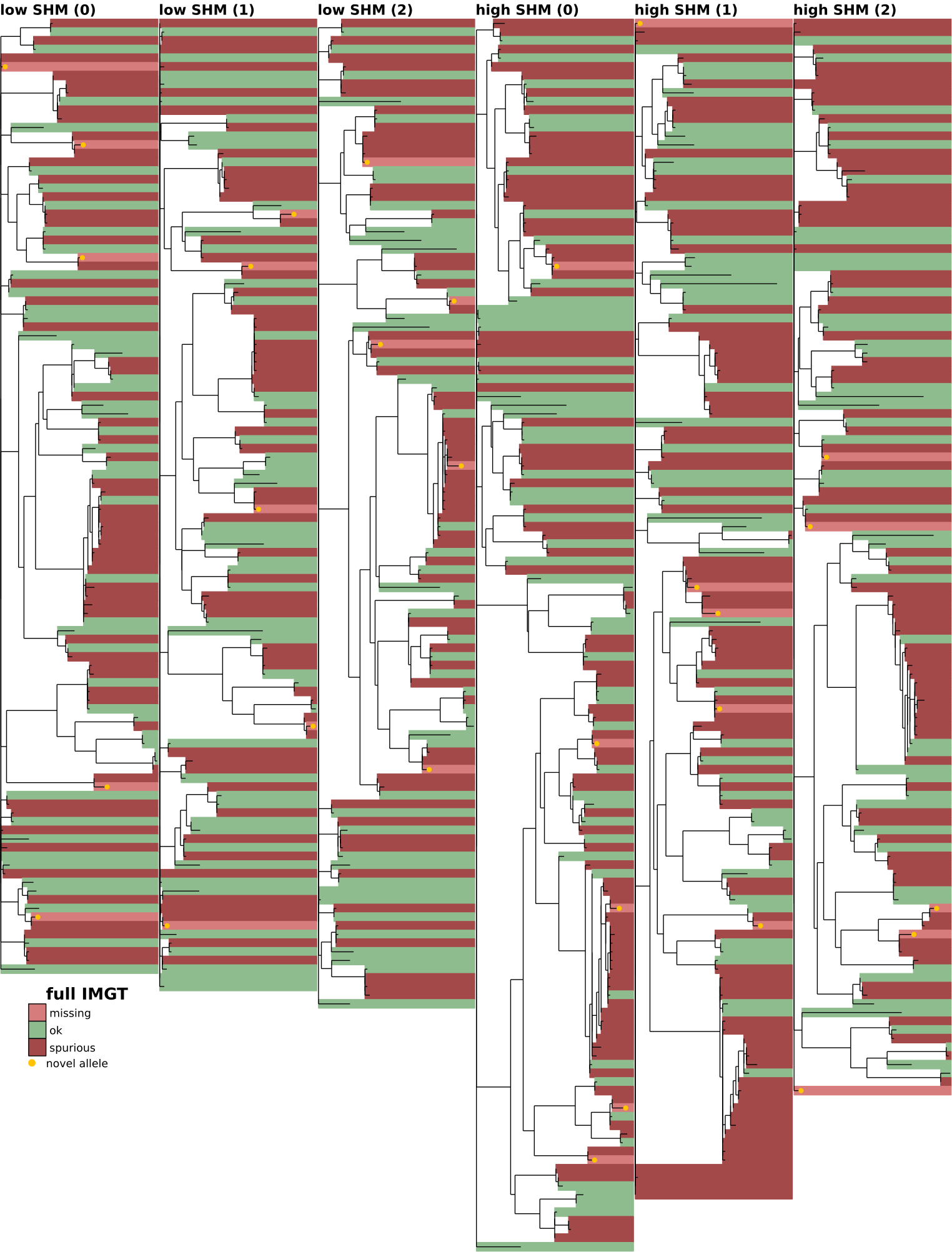}}
\caption{\
  {\bf Full-repertoire germline set accuracy for the currently widespread method of aligning every sequence to its closest match in the full \imgt\ V gene set}.
  The phylogenetic tree is constructed with a leaf for each germline gene in either the true or inferred germline sets (see Methods).
  \blBlurb\
  Leaves are colored according to the similarity of the true and inferred germline sets, with shared genes in green and unshared in red, the latter broken into missing (light red) and spurious (dark red).
  Novel alleles (not in the \imgt\ database, whether from the true simulated set or spuriously inferred) are highlighted in gold.
  \glssamples\
}
\label{FIGglsSimFull}
\end{figure}

\begin{figure}[!ht]
\forarxiv{\includegraphics[width=3in]{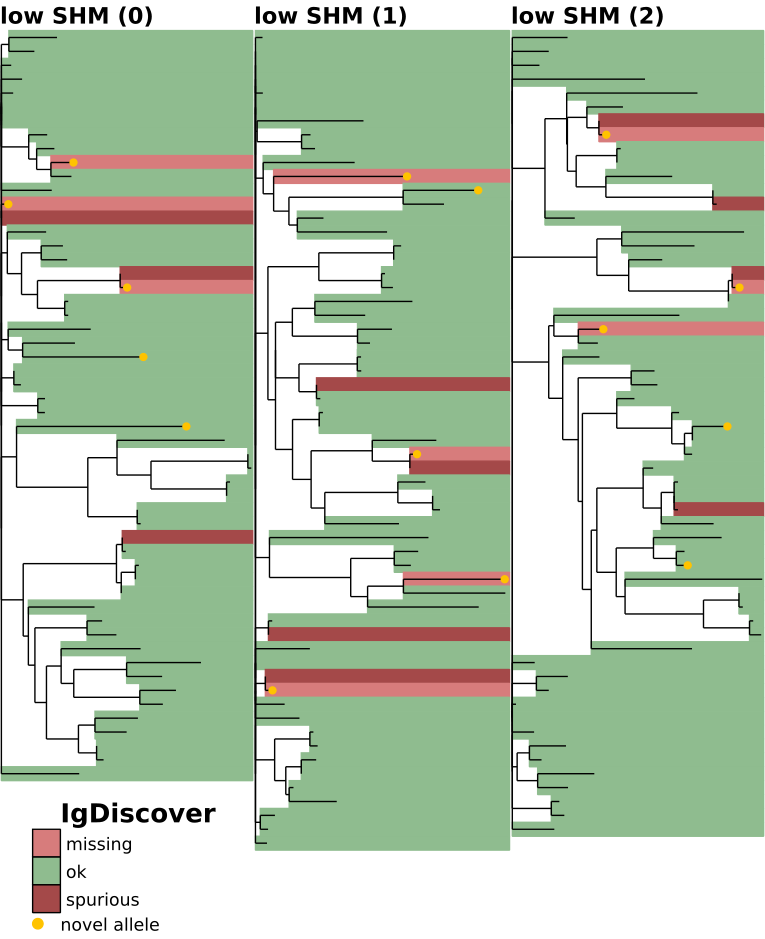}}
\caption{\
  {\bf Full-repertoire germline set accuracy for \igdiscover} (explanation in Fig~\ref{FIGglsSimFull}).
  Shown on the first three replicates (0-2) of the low-SHM full-repertoire simulation samples. The high-SHM samples are not shown, since \igdiscover\ is designed only for low-SHM IgM samples (see text).
}
\label{FIGglsSimIgdiscover}
\end{figure}

\begin{figure}[!ht]
\forarxiv{\includegraphics[width=5in]{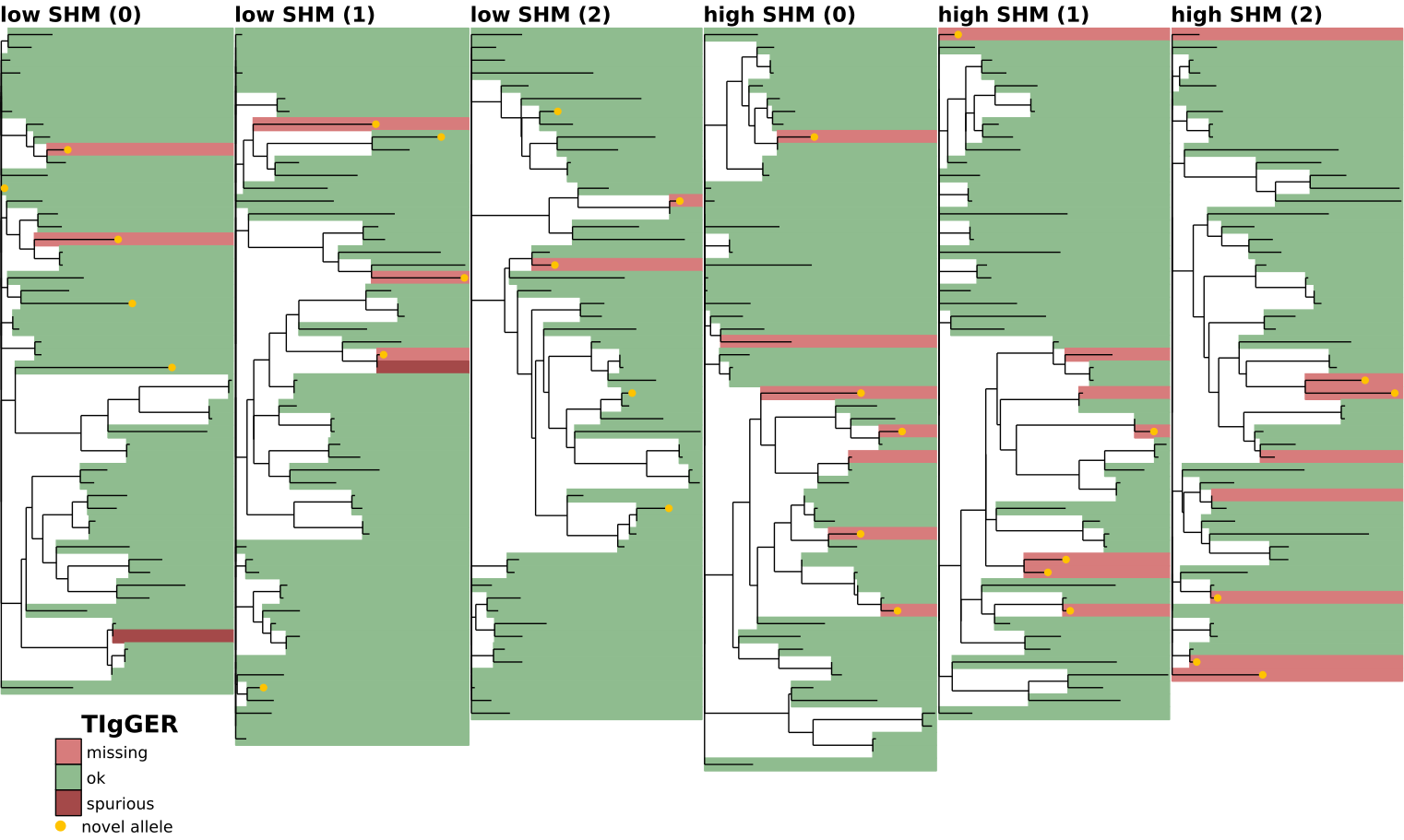}}
\caption{\
  {\bf Full-repertoire germline set accuracy for \tigger} (explanation in Fig~\ref{FIGglsSimFull}).
}
\label{FIGglsSimTigger}
\end{figure}

\begin{figure}[!ht]
\forarxiv{\includegraphics[width=5in]{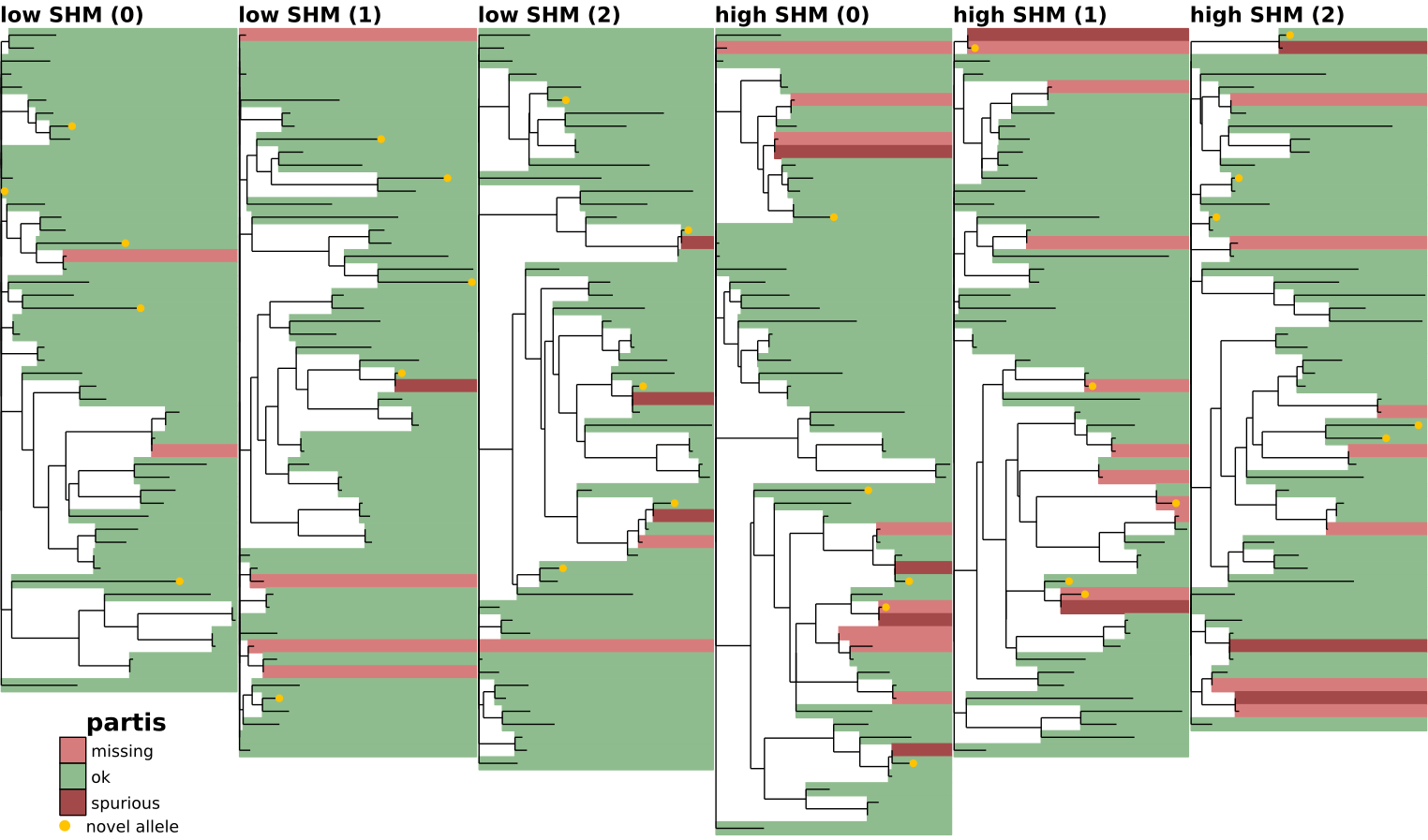}}
\caption{\
  {\bf Full-repertoire germline set accuracy for partis} (explanation in Fig~\ref{FIGglsSimFull}).
}
\label{FIGglsSimPartis}
\end{figure}

\newcommand{\nonIgmDataBlurb}{Since this data is not IgM specific, \igdiscover\ is not shown. }
\newcommand{\dataTreeBlurb}{
  \blBlurb\
}
\newcommand{\treeLeafFragment}{The phylogenetic tree is constructed with a leaf for each germline gene that was inferred}
\newcommand{\leafnameblurb}[1]{The same trees, but with leaves labeled with gene names, are shown in Fig~\ref{#1}.}

\begin{figure}[!ht]
\forarxiv{\includegraphics[width=5in]{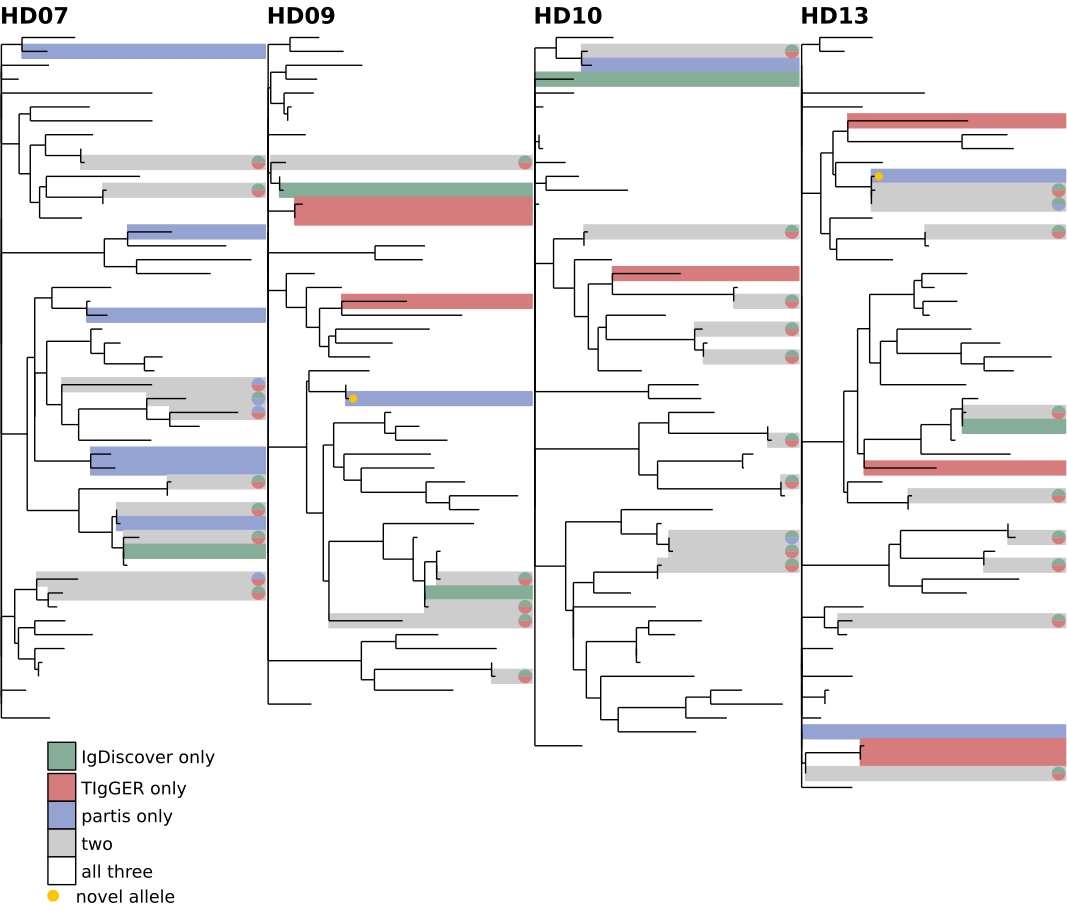}}
\caption{\
  {\bf Comparison of all three inference methods on the healthy donor samples from~\cite{Vander_Heiden2017-fs}} (other subjects shown in Figs~\ref{FIGdataJasonMGARmethodVsMethod} and~\ref{FIGdataJasonMGMKmethodVsMethod}).
  \treeLeafFragment\ by any of the methods.
  \dataTreeBlurb\
  Leaves are colored according to how many methods inferred the corresponding gene: one (green, red, blue), two (grey), or all three (white).
  \leafnameblurb{FIGdataJasonMGHDmethodVsMethodLeafNames}
}
\label{FIGdataJasonMGHDmethodVsMethod}
\end{figure}

\begin{figure}[!ht]
\forarxiv{\includegraphics[width=3in]{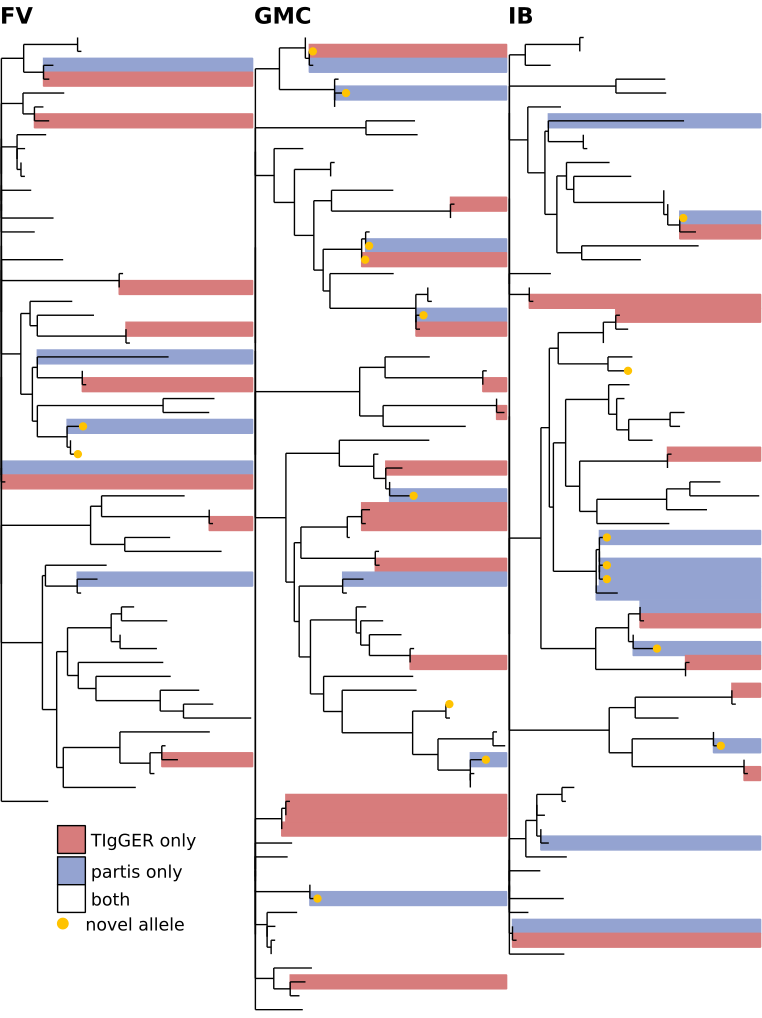}}
\caption{\
  {\bf Comparison of germline sets inferred by partis and \tigger\ for subjects FV, GMC, and IB from~\cite{Laserson2014-yh}}, with all ten time points merged for each subject.
  \treeLeafFragment\ by either of the two methods.
  \dataTreeBlurb\
  Leaves are colored according to how many methods inferred the corresponding gene: either one (red, blue) or both (white).
  \nonIgmDataBlurb\
  Includes the three time points in Fig~\ref{FIGdataJasonInfluenzaSampleVsSample}, plus seven more, for each subject.
  \leafnameblurb{FIGdataJasonInfluenzaMethodVsMethodLeafNames}
}
\label{FIGdataJasonInfluenzaMethodVsMethod}
\end{figure}

\begin{figure}[!ht]
\forarxiv{\includegraphics[width=4in]{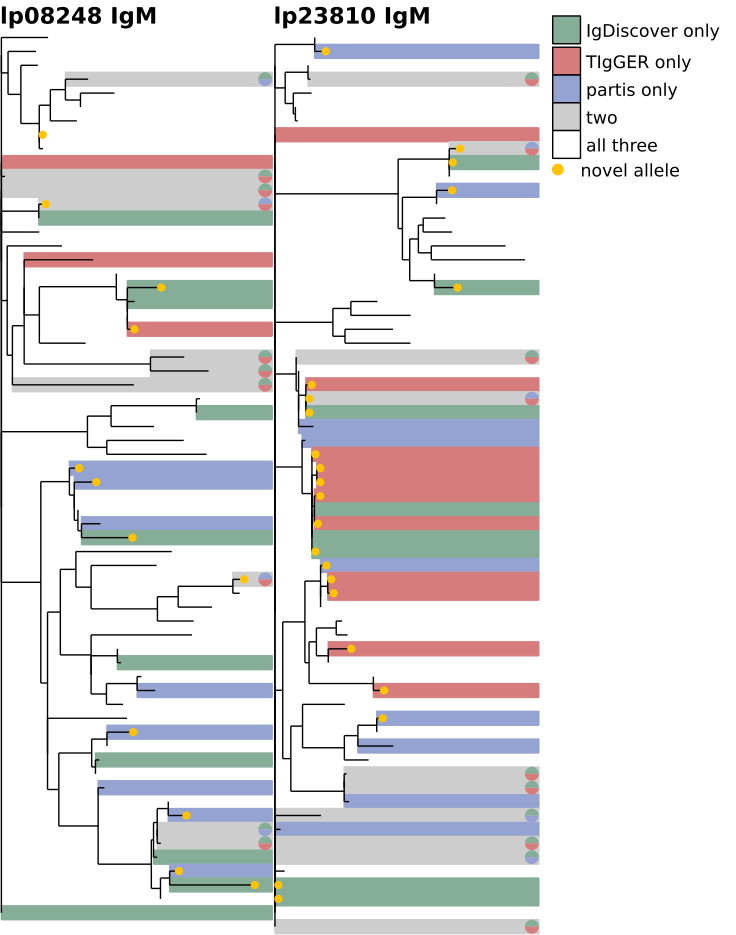}}
\caption{\
  {\bf Comparison of the three methods on IgM data} from subjects lp08248 (left) and lp23810 (right) from~\cite{Sheng_Schramm_Kong_NISC}.
  \treeLeafFragment\ by any of the methods.
  \dataTreeBlurb\
  Leaves are colored according to how many methods inferred the corresponding gene: one (green, red, blue), two (grey), or all three (white).
  See Fig~\ref{FIGdataShengGSSPsampleVsSample} for other results for these subjects.
  \leafnameblurb{FIGdataShengGSSPMethodVsMethodLeafNames}
}
\label{FIGdataShengGSSPMethodVsMethod}
\end{figure}

\begin{figure}[!ht]
\forarxiv{\includegraphics[width=4in]{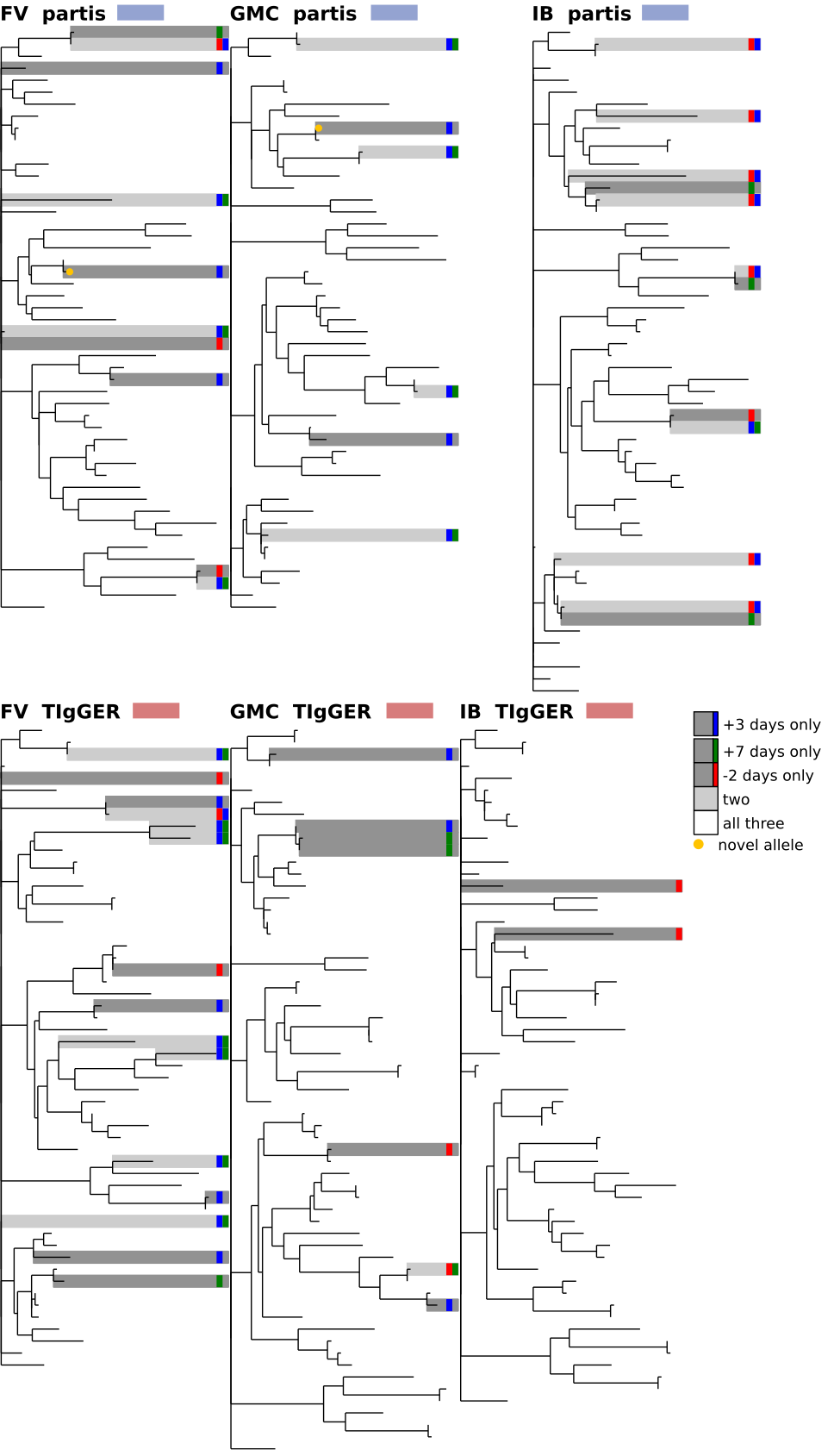}}
\caption{\
  {\bf Comparison of inferred germline sets for samples taken at different time points for subjects FV, GMC, and IB from~\cite{Laserson2014-yh}.}
  Shown for three (of ten total) time points surrounding influenza vaccination: two days before, three days after, and seven days after; for partis (top) and \tigger\ (bottom).
  \treeLeafFragment\ at any of the three time points.
  \dataTreeBlurb\
  Leaves are colored according to the number of time points at which the corresponding gene was inferred: one (dark grey), two (light grey), or all three (white).
  \nonIgmDataBlurb\
  See Fig~\ref{FIGdataJasonInfluenzaMethodVsMethod} for other results for these subjects.
  \leafnameblurb{FIGdataJasonInfluenzaSampleVsSampleLeafNames}
}
\label{FIGdataJasonInfluenzaSampleVsSample}
\end{figure}

\begin{figure}[!ht]
\forarxiv{\includegraphics[width=5in]{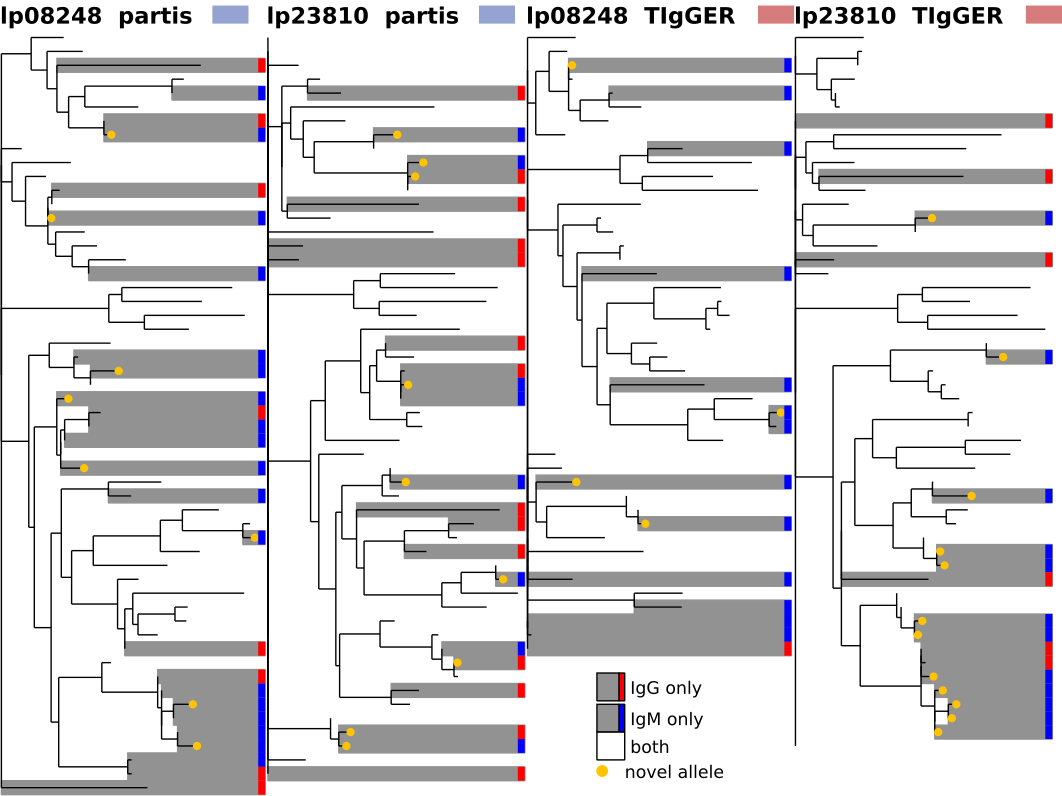}}
\caption{\
  {\bf Comparison of inferred germline sets for IgM vs IgG data} from subjects lp08248 and lp23810 from~\cite{Sheng_Schramm_Kong_NISC} for partis (left) and \tigger\ (right).
  \treeLeafFragment\ for either of the two isotypes.
  \dataTreeBlurb\
  Leaves are colored according to the number of isotype-specific samples for which the corresponding gene was inferred:  either one (grey) or both (white).
  See Fig~\ref{FIGdataShengGSSPMethodVsMethod} for other results for these subjects.
  \leafnameblurb{FIGdataShengGSSPsampleVsSampleLeafNames}
}
\label{FIGdataShengGSSPsampleVsSample}
\end{figure}

\begin{figure}[!ht]
\forarxiv{\includegraphics[width=5in]{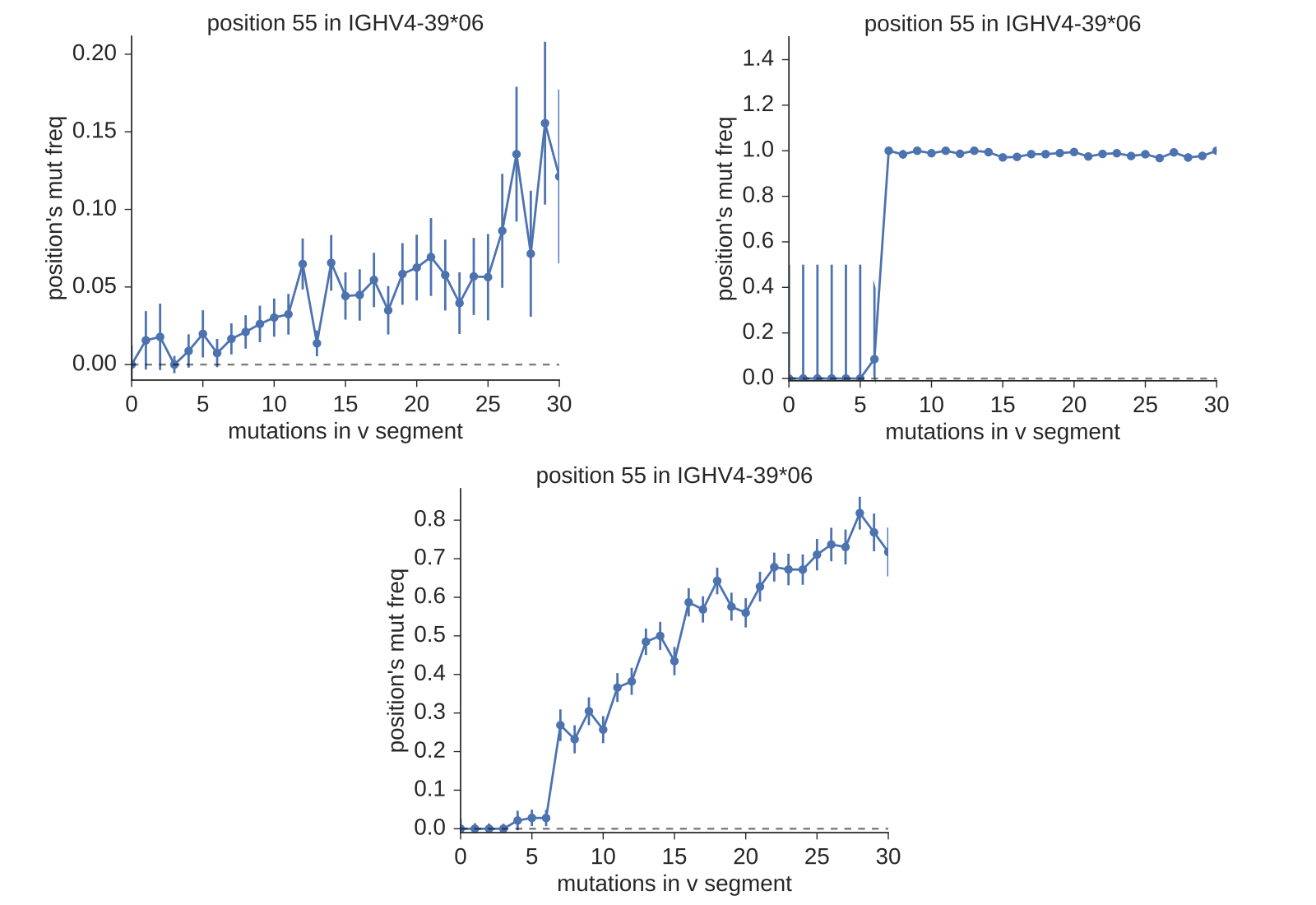}}
\caption{\
  Mutation accumulation plots showing the relationship between the mutation probability at position 55 across all sequences aligning closest to IGHV4-39*06 (y-axis), and the number of mutations in the entire observed V sequence (x-axis) for three simple, hypothetical BCR repertoires.
  In the top row are two repertoires that consist of a single allele: where this allele is known (left), and where it is unknown, but separated by seven SNPs from a known allele (right).
  In a more typical case, given the relative completeness of the standard germline sets, we would observe a mixture of sequences from the known and unknown alleles (bottom).
  This is equivalent to the (shifted) superposition of the two plots in the top row.
}
\label{FIGbasicIdea}
\end{figure}

\begin{figure}[!ht]
\forarxiv{\includegraphics[width=5in]{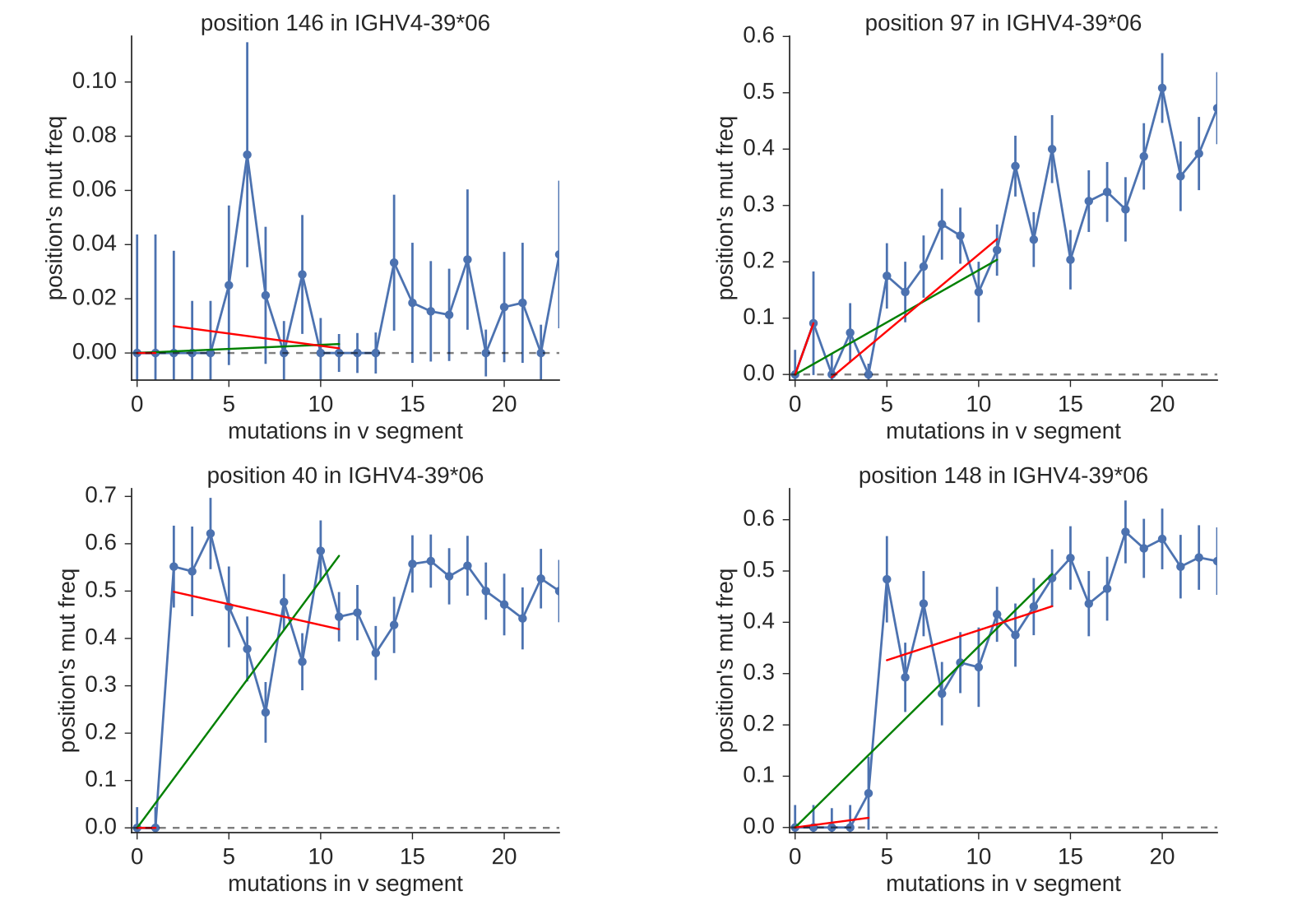}}
\caption{\
  {\bf Example one-piece (green) and two-piece (red) fits for positions without (top row) and with (bottom row) evidence for new alleles.}
  The left and right plots in the top row show the difference between positions with low and high mutability (cold and hot spots).
  The bottom row shows a position with evidence for a new allele with \nsnp\ equal to two (left) and a similar plot for \nsnp\ equals five (right).
  Note that both one-piece and two-piece models fit well in the top row, whereas in the bottom row only the two-piece model provides an adequate fit.
}
\label{FIGsimpleFits}
\end{figure}

\subsubsection*{Variation of individual variables on sparse repertoires}
Using partis's germline set inference algorithm, we quantified the impact of six repertoire characteristics on sensitivity and specificity.
We did so by plotting the fraction of alleles in the true repertoire that are missing from the inferred repertoire, and the fraction of spuriously-inferred alleles (that are not in the true repertoire), as a function of sample size for each variable (\varvalfigs).

Increasing the rate of SHM makes inference more challenging (\varvalfigs, top left, with the corresponding SHM distributions in Fig~\ref{FIGvFreqDistr}).
Because allele inference sensitivity is determined mainly by sequences with a small number of SHMs (specifically, a number comparable to the number of SNPs separating the new and existing alleles), raising SHM rates effectively reduces sample size.

Alleles that occur at low prevalence are more difficult to infer: as the fraction of sequences stemming from the new allele decreases, so does sensitivity (\varvalfigs, top right).

The number of SNPs (\nsnp) separating a new allele from its most similar known counterpart also affects the details of germline inference.
We show performance for different \nsnp\ for both a single new allele (\varvalfigs, middle left) and for several combinations of multiple new alleles (\varvalfigs, middle right).
Sensitivity is independent of \nsnp\ for smaller \nsnp\ (three or less), and then decreases slightly with increasing \nsnp.
The presence of multiple new alleles, on the other hand, does not appreciably affect sensitivity as long as their SNPs do not occur at the same positions.
Because the occurrence of multiple new alleles with the same SNP positions is rare in real data, we do not show results for this case.
In many cases it is in fact possible to disentangle such alleles, but this depends on the details of each new allele's prevalence and \nsnp.

The shared mutations within a clonal family complicate allele inference because independent mutations are required for accurate fitting (see Methods).
We find that increasing clonality effectively decreases sample size (\varvalfigs, bottom left), rather than introducing the spurious alleles that would result from fitting with non-independent mutations.
This indicates that our method of selecting a small number of sequences to represent each clonal family (see Methods) provides a sufficiently accurate method of choosing sequences with independent mutations.

We find that variations in phylogenetic tree shape do not greatly affect our method (\varvalfigs, bottom right).
We change tree shape by using the TreeSimGM package~\cite{Hagen2015-ob} to vary the shape parameter of a Weibull distribution controlling an age-dependent speciation process.  

These single-variable results show that our method's sensitivity is high enough to give useful results with the sample sizes and SHM rates characteristic of typical full-repertoire samples, and that it models repertoire details well enough that spurious alleles are rare.
Note that \tigger\ and \igdiscover\ are not shown on these sparse samples because both methods use hard-coded assumptions tailored to typical full repertoires that cause crashes on these sparse repertoires.

\subsubsection*{Full-repertoire samples}
In the second validation stage, we show performance on a smaller number of large, realistic repertoires using partis (v0.9.0), \tigger\ (v0.2.10), \igdiscover\ (v0.6.0), and annotation with the full \imgt\ set.
All software was run with default parameters.

We split these full-repertoire samples among two difficulty levels: ten samples with more-uniform allele prevalence and low SHM, and ten samples with less-uniform allele prevalence and higher, typical SHM (details in Methods).
All of these samples contain 50,000 sequences.
Results with \igdiscover\ are shown only for the low-SHM samples, since \igdiscover\ is designed to work only on low-SHM IgM-specific data.
We note that the \igdiscover\ manual recommends sample sizes of at least 750,000 sequences; it is unclear what this recommendation is based on, as we do not observe \igdiscover\ to have any sensitivity issues on much smaller samples.

We measure the influence of germline set accuracy on practical results in two ways: in terms of the actual genes and alleles inferred, and in terms of the resulting annotation accuracy.
The former is more relevant to germline databases and studies of gene association, while the latter is of more concern when inferring and studying the function of ancestral sequences.

We find that the practice of aligning against the full \imgt\ set results in a very large number of spurious gene inferences, even on low-SHM samples (Table~\ref{TABLEglsSimSummary} and Fig~\ref{FIGglsSimFull}).
The three explicit germline inference methods, while all giving much smaller numbers of spurious genes, harbor significant differences.
The partis-inferred missing and spurious alleles are found on relatively short branches compared to those of the other programs (Figs~\ref{FIGglsSimIgdiscover},~\ref{FIGglsSimTigger},~\ref{FIGglsSimPartis}).
This results in partis's significantly more accurate V naive inference (Fig~\ref{FIGglsSimAnnotationPerformance}).
By considering the distribution of Hamming distances between true and inferred naive V sequences (Fig~\ref{FIGglsSimAnnotationPerformance}), we see that the relative inaccuracy of \tigger\ and \igdiscover\ is driven by rare sequences that are assigned to genes that are very dissimilar to their true gene.
We also note that \tigger\ shows reduced sensitivity at typical SHM rates (Fig~\ref{FIGglsSimTigger} right), compared to low SHM rates (Fig~\ref{FIGglsSimTigger} left), in fact failing to infer any of the novel (non-\imgt) alleles at typical SHM rates.

These simulation samples, together with true and inferred germline sets, are available at \zenodolink.

\subsection*{Results on real data}
In order to evaluate performance on real data, it would be natural to deep sequence individuals for whom we also have accurate results from germline sequencing.
Unfortunately, as described above, the difficulty of germline sequencing means that such samples are not readily available.
We instead use two types of comparison that, while not definitive, provide some insight.

\begin{figure}[!ht]
\forarxiv{\includegraphics[width=5in]{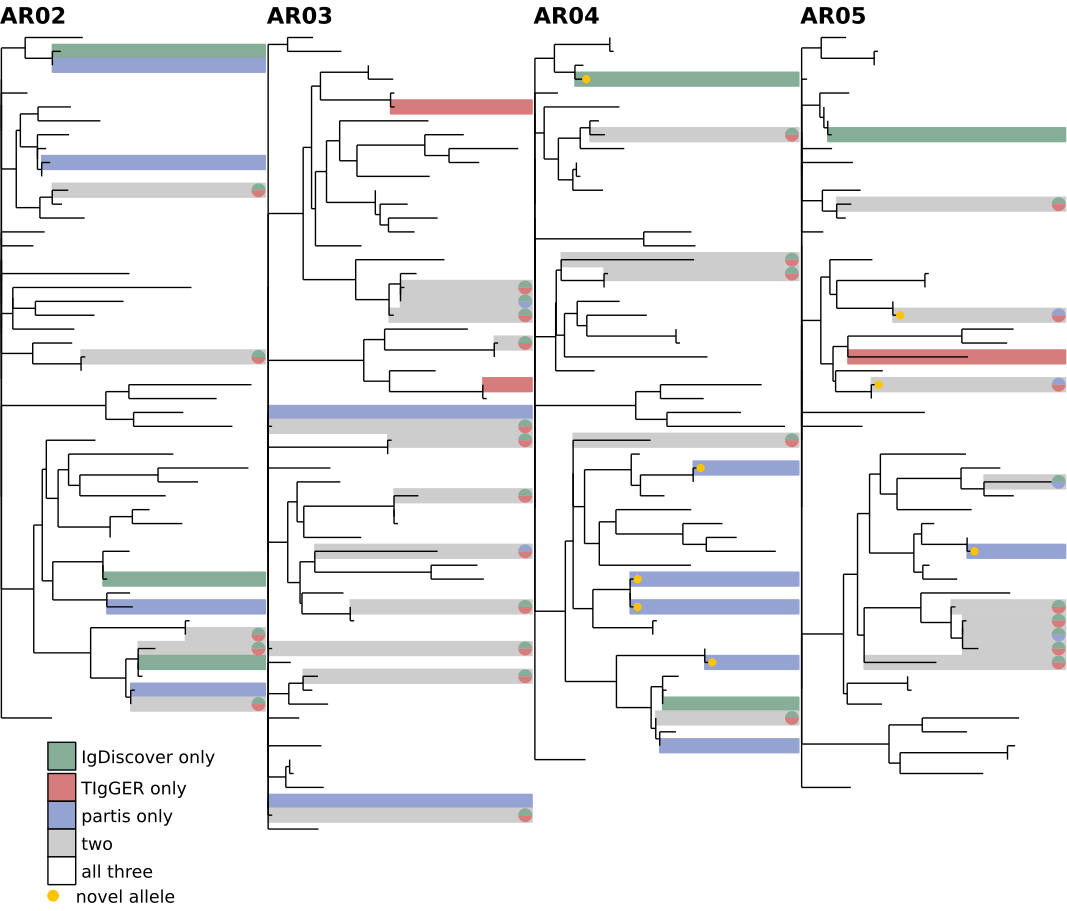}}
\caption{\
  {\bf Comparison of all three inference methods on the AR-type Myasthenia Gravis samples from~\cite{Vander_Heiden2017-fs}} (explanation in Figure~\ref{FIGdataJasonMGHDmethodVsMethod}).
}
\label{FIGdataJasonMGARmethodVsMethod}
\end{figure}

\begin{figure}[!ht]
\forarxiv{\includegraphics[width=5in]{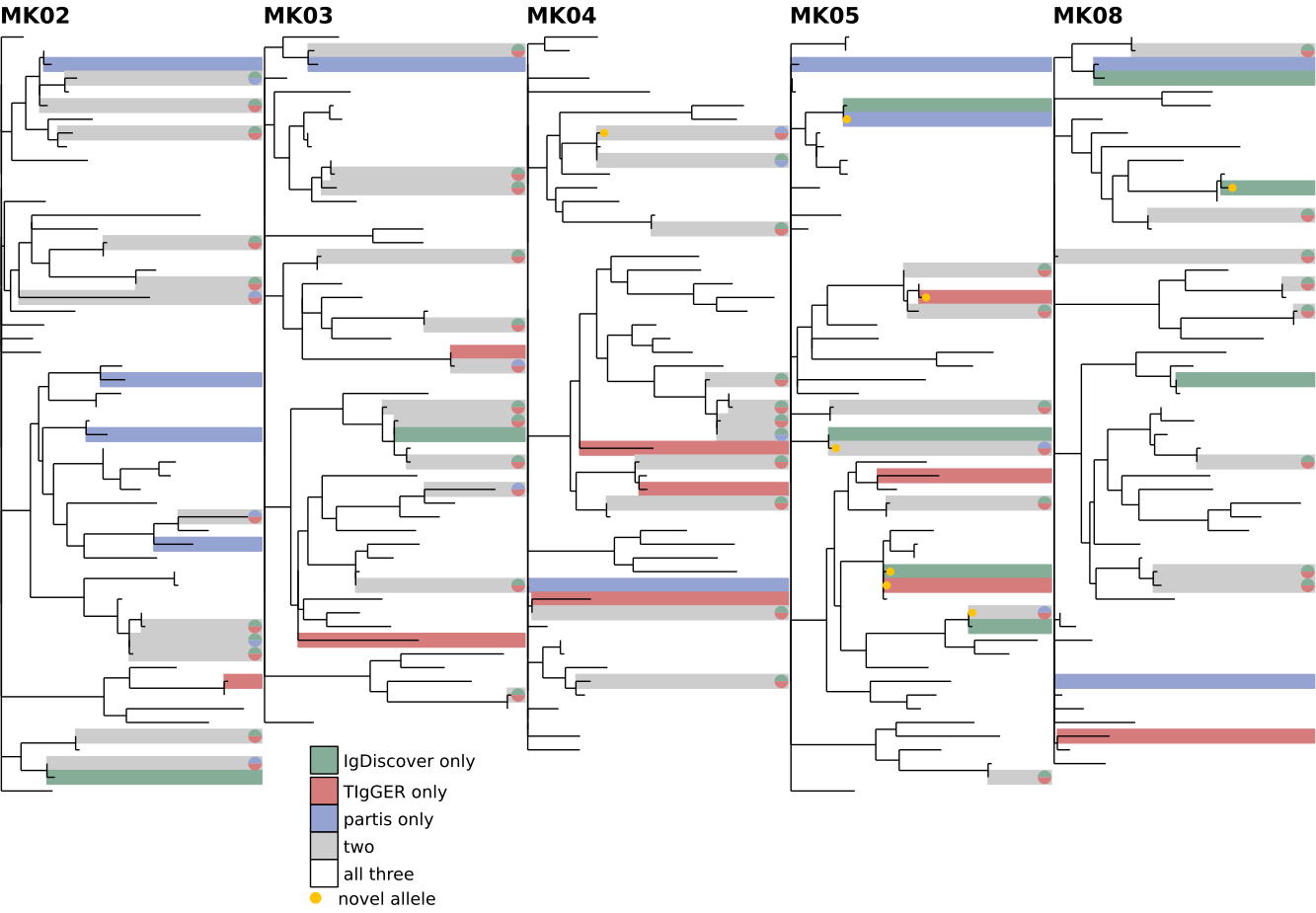}}
\caption{\
  {\bf Comparison of all three inference methods on the MK-type Myasthenia Gravis samples from~\cite{Vander_Heiden2017-fs}} (explanation in Figure~\ref{FIGdataJasonMGHDmethodVsMethod}).
}
\label{FIGdataJasonMGMKmethodVsMethod}
\end{figure}

We first compare results from the different inference methods when run on the same sample, and find agreement on 70-90\% of the total genes (Figs~\methodVsMethodFigures).
While this is reassuring, some caution is advised, as the methods are far from uncorrelated (see Supplement).
\igdiscover\ is shown only for IgM samples (Fig~\ref{FIGdataShengGSSPMethodVsMethod}, and non-IgM samples with very low SHM rates (Figs~\ref{FIGdataJasonMGHDmethodVsMethod},~\ref{FIGdataJasonMGARmethodVsMethod},~\ref{FIGdataJasonMGMKmethodVsMethod}).
Also of note is the large cluster of closely-related novel alleles inferred only by \tigger\ in the IgM data from subject lp23810~\cite{Sheng_Schramm_Kong_NISC} (Figs~\ref{FIGdataShengGSSPMethodVsMethod},~\ref{FIGdataShengGSSPsampleVsSample}).

We next compare the results of each inference method on several different samples from the same individual. 
We find a similar overall level of agreement both when comparing samples from different time points (Fig~\ref{FIGdataJasonInfluenzaSampleVsSample}), and of different isotypes (Fig~\ref{FIGdataShengGSSPsampleVsSample}).
These comparisons give some idea of each method's uncertainty because, while the physical germline genes are in each case identical, the SHM rates, gene expression levels, and clonal family structure vary significantly with both time and isotype.

We also use the partis-inferred germline sets to make an estimate of the number of genes that are expressed at levels too low for us to detect.
Previous work has reported a range of values for the total number of functional V genes per individual.
One study~\cite{Watson2013-heavy} reported 43 full-length functional V genes (plus 1 truncated) for a single haplotype, while another~\cite{lefranc-fact-book} reported a range of 38-46 per haplotype.
In order to convert these per-haplotype totals to per-diplotype totals, we calculate the mean fraction of alleles shared between the inferred germline sets from two unrelated individuals.
For the sequencing data in this paper, this mean overlap is 67\% (range 50-85\%).
This suggests that to go from per-haplotype to per-diplotype totals we multiply by $1 + (1 - 0.67)$, which yields per-diplotype estimates of 57 for~\cite{Watson2013-heavy} and 51-61 for~\cite{lefranc-fact-book}.
These values, both for total genes and for the fraction of genes shared between unrelated parental haplotypes, roughly agree with two other studies that found 35-46 per haplotype and 39-55 per diplotype~\cite{kidd2012-cj}, and 45-60 per diplotype with a mean alleles per gene of 1.2~\cite{boyd2010-yj}.
The mean total number of partis-inferred V genes observed in individuals in this paper, meanwhile, is 47 (range 38-62).
This suggests that the sample sizes, clonal family structures, mutation rates, and expression levels, together with our method's sensitivity, result in a failure to detect about 0 to 10 genes per individual.
We have not accounted for spuriously-inferred alleles in this calculation because our validation results suggest that when partis does infer spurious alleles, each simply replaces a very similar true allele, and thus does not have an appreciable net effect.

A summary of the data samples used is shown in Table~\ref{TABLEdataSamples}.
The fasta files for each inferred germline set are available at \zenodolink.
We have made the command-line script used to make phylogenetic comparison plots available for general application at \url{https://git.io/vFo2B}.

\section*{Discussion}
We have developed a practical new tool for inferring per-sample immunoglobulin germline gene sets, and performed extensive validation and comparison against existing tools.
Our tool is implemented in the existing partis annotation and clonal family inference package.
We have shown that in simulation the currently widespread practice of aligning expressed BCR samples against the full \imgt\ germline set results in both large numbers of spurious alleles and inaccurately inferred naive ancestors.
We next showed that on simulation our method infers significantly more accurate germline sets than the existing \tigger\ and \igdiscover\ methods in terms of both inferred gene similarity and naive ancestor inference, but of similar accuracy in terms of raw number of genes.
We then compared germline sets inferred by the three methods on a variety of real data samples, which showed generally similar features to those in simulation.
Together, these results comprise the first comparison of any germline inference methods, and thus provide users with a valuable baseline for expected accuracy, while also demonstrating the inaccuracy of the widespread method of aligning against the entire \imgt\ set.

While our method has reached a level of maturity such that it is now run by default in the partis annotation and clonal family inference procedures, it has a number of weaknesses.
First, as with the rest of partis, it assumes that all corrections for sequencing error have been performed before input.
Second, our piecewise-linear model for the mutation accumulation plots (see Methods) is only an approximation of the real behavior.
Thus, while we have designed our method with the aim of maximizing robustness against atypical repertoires, a more complex model that more closely modeled the repertoire's nonlinearities would provide better performance.
Another source for improved performance would be the incorporation of per-base mutation information, i.e. splitting apart the mutation accumulation plots by A, C, G, and T.
Additionally, because we do not yet set any prior on the number of germline V genes, our method will underestimate this number on smaller samples (roughly a few thousand sequences or less).
Also, we have thus far only applied our method to V region genes, although the extension to D and J should be conceptually straightforward.
We also at this point only report the most likely germline set, with no probabilistic guidance as to the likelihood of other possible germline sets (although some idea of uncertainty can be gleaned from the fits and goodness of fit metrics).
This is an unfortunate consequence of the intricacies of the optimizations necessary to make the method usefully fast, together with the lack of good information about the variation of the Ig locus between individuals.
Finally, taking advantage of the fact that rearrangement occurs only between genes on the same chromosome, as in~\cite{kidd2012-cj,Elhanati2015-ld}, would likely provide additional improvement.

A further limitation of our method is that it looks only for new alleles separated by SNPs from existing alleles, and not for those separated by insertion/deletion events.
While this is not a significant limitation on human samples, the \imgt\ germline sets for other species are incomplete enough that, in those species, this could cause novel alleles to be misinterpreted as SHM indels.
This is one respect in which the clustering-based approach taken by \igdiscover\ offers a significant advantage (see Supplement).  
For this reason we have also implemented a non-default clustering-based method which can be run in addition to the purely mutation accumulation plot-based method described here (see Manual).
While we thus recommend this clustering-based method for non-human samples, its robustness, like \igdiscover's, can suffer on some highly-mutated samples, so we have left it as a non-default option pending future improvements.

\section*{Methods}
\subsection*{Overview}
The task of inferring germline genes consists largely of learning to distinguish between positions that are highly mutated as a result of SHM, and those whose highly-mutated appearance stems from the occurrence of previously unknown alleles.
A few key observations allow us to extract enough information to make this distinction.
First, in the absence of unknown alleles, the probability of a mutation at each position in an observed sequence is roughly proportional to the total number of mutations in that sequence (at least at the low SHM levels relevant for new-allele inference).
In other words, while mutation rates differ dramatically from position to position according to, for instance, hot and cold spot motifs, each position is more likely to be mutated in sequences that have been subject to higher levels of SHM.
Almost all of the power to infer new germline sequences comes from the low-SHM sequences in the sample, and for those sequences the hotness or coldness of most positions will not change upon introduction of the first few mutations.
In the presence of unknown alleles, on the other hand, sequences stemming from these unknown alleles will be mistakenly assigned to the most similar known allele, causing this approximate proportionality to be violated.
If there are, say, \nsnp\ SNPs separating a known and unknown allele, then there will be very few sequences from this unknown allele that appear to have fewer than \nsnp\ mutations.
The \nsnp\ positions at which they differ, on the other hand, will almost always appear to be mutated in sequences that appear to contain \nsnp\ or more total mutations.
This differing apparent mutational behavior between sequences with fewer than, as compared to more than, \nsnp\ mutations provides the basis for our method.

A convenient way to visualize these observations is with a type of plot introduced in~\cite{tigger}, which we call a ``mutation accumulation'' plot.
To make a set of these plots for one germline gene, we first collect every sequence in the sample that aligns most closely to this single known gene.
We then use these sequences to make one plot for each nucleotide position as follows.
The sequences are binned along the x-axis according to their total number of apparent V mutations.
The y-coordinate of each bin, meanwhile, is the frequency with which that plot's nucleotide position appears to mutate among the sequences in the bin.
For the full repertoire, we first group sequences based on their closest known germline gene, and then follow the procedure above for each such group.
We first show example plots for three simple, hypothetical repertoires (Fig~\ref{FIGbasicIdea}).
While these simple repertoires, by themselves, are gross simplifications of the biological complexity in a real BCR repertoire, they contain the essential elements from which we can construct a method that performs well on real data sets.

\subsection*{Models and fitting}
\newcommand{\rr}{\ensuremath{r}}

In the context of mutation accumulation plots (Fig~\ref{FIGbasicIdea}), the presence of new alleles is signaled by a departure from what would be expected if all sequences had been assigned to the correct true gene.
Namely, to the extent that mutations at each site accumulate in proportion to the total number of mutations in the sequence, correct assignment would result in simple linearity.
For incorrect assignment, this linearity is replaced with differing behavior between the regions below and above \nsnp.
Our task, then, amounts to distinguishing between plots that can be adequately described by a one-piece linear model, and those that require a model consisting of two pieces separated by a discontinuity.

In order to distinguish these two hypotheses, we construct a model for each.
The one-piece model is simply a linear fit constrained to pass through the origin.
The two-piece model, meanwhile, consists of two separate linear fits, which we call the ``lower'' (below \nsnp) and ``upper'' (above \nsnp) fits.
The lower fit is constrained to pass through the origin, while the upper fit's y-intercept must be near the average of the upper-region mutation frequencies (within 1.5 standard deviations of their mean).
The junction between the two pieces must harbor a significant discontinuity in either bin value (mutation frequency) or bin total (number of sequences per bin), where significance is defined as a difference of more than 2.5 times the larger uncertainty.
This two-piece model describes the presence of a new allele separated by \nsnp\ SNPs from the original known gene.
To give a general idea of the implementation, several examples are shown in Fig~\ref{FIGsimpleFits}.

We use a ratio of error descriptors to determine whether a plot is adequately described by the one-piece fit.
Define $\epsilon$ to be the sum of squared residuals divided by degrees of freedom, which in regression analysis is sometimes called the mean squared error.
Good fits are characterized by values of $\epsilon$ around one, while values much greater than one indicate poor fits.
Values significantly less than one generally indicate poorly-estimated uncertainties.
For our purposes, then, we are interested in positions (which we call ``candidate positions'') for which $\epsilon$ is large for the one-piece fit (greater than 4.5) but around one (less than 1.95) for the two-piece fit.

For each \nsnp, we construct the most plausible potential new allele by finding the \nsnp\ positions that have the worst one-piece, but best two-piece, fits.
We quantify this using the ratio of the two $\epsilon$,
\begin{equation}
  \rr = \frac{\epsilon_{\text{1-piece}}}{\epsilon_{\text{2-piece}}}.
\end{equation}
Because cases that would be better described by more complex models will have larger residuals (poor fits) for both one-piece and two-piece models, which cancel out in the ratio, this formulation provides robustness to deviations from linearity.
The model for the best potential new allele consists of the \nsnp\ positions that have the largest values of \rr.

In order to strike an appropriate balance between focusing the fit's attention on the area of the discontinuity, while taking advantage of the largest possible sample size from many surrounding bins, we perform all fits in a window of width 10 bins.
This window begins at zero for small \nsnp, while for larger \nsnp\ it is symmetric around \nsnp.


We apply several additional criteria to ensure that the candidate fits make a compelling case for a new allele.
The slope at the discontinuity, i.e.\ the slope defined by the two points on either side, must be much larger than both the upper and lower fitted slopes (a fractional difference of more than 2.5 times).
For larger \nsnp\ (five or more), the slopes before and after the discontinuity must also either be consistent, or the lower slope must be the smaller of the two.

The unfortunate profusion of constant values in the preceding paragraphs deserves some examination.
In general, for the sake of simplicity and interpretability, we have wherever possible minimized the number of such constants.
However, practical constraints make it difficult to reduce their number further.
In theory, it would be possible to construct a more complicated model that faithfully recreated all the details of the real system, which would enable a collection of simple likelihood ratio tests.
However, in practice this approach is unlikely to be computationally feasible, and would likely require a much lengthier development process.
Instead, we have adopted the approach of comprehensively validating a simpler model which, nevertheless, provides an adequate description of the system's real biological complexity.

\subsection*{Comparing multiple hypotheses}
The previous section outlines a procedure for identifying a single potential new allele for each individual \nsnp.
In realistic samples, however, we must treat the general case where there may be several new alleles, either with the same \nsnp, or spread among several \nsnp.

To do this, we first sort every candidate position within each \nsnp\ by decreasing \rr.
In order to better adjudicate between ties in the first sort, we then sort again either by decreasing y-intercept (if \nsnp\ less than three) or decreasing two-piece fit $\epsilon$.
The first \nsnp\ elements of this sorted list of candidate positions are then taken as a candidate allele, the next \nsnp\ positions are taken as a second candidate allele, and so on, until fewer than \nsnp\ remain.
The second sorting step serves to group together positions with similar fit properties, and that are thus most likely to come from the same new allele.
These fit properties are affected by several aspects of the new alleles, most notably their prevalence.
In cases with two new alleles with the same prevalence, for example, this is not an effective means of determining which positions go with which allele; however, in real data such cases are very rare.

For each of these candidate alleles, both the smallest \rr\ among their positions, and the mean, must be greater than 2.75.
The discontinuities for every pair of positions must also be compatible, defined as the difference in bin totals (number of sequences) on either side of \nsnp, which must be closer than three times the maximum of their two uncertainties.

This procedure is repeated for each \nsnp, resulting in a list of candidate alleles from each; these lists are then merged into a final list that is sorted by decreasing \nsnp.
We then go through this list and discard alleles that share any positions with an allele earlier in the list.
This last sorting is due to the fact that it is easier for a high-\nsnp\ allele to mimic a low-\nsnp\ allele than the reverse.


\subsection*{Approximations and pre-filters}

The procedure described above would work well in principle, but would require a computationally prohibitive number of fits.
As a rough estimate, taking 50 initial, known alleles in a sample, each with 300 positions, looking up through \nsnp\ equals eight, and with both the one-piece and two-piece fits, we would need 360,000 individual linear fits.
To be useful, however, it must run as part of the overall partis annotation, which takes only minutes on samples with tens of thousands of sequences.
Luckily, the overwhelming majority of these fits can be avoided by ignoring uninteresting positions using a number of approximation procedures.
The cumulative effect of the following approximations and filters is that a typical run requires of order 100 fits, with no appreciable decrease in precision or sensitivity.
This results in a method that does not add significant run time to an existing partis run.

The first step is to ignore positions for which there would not be enough statistical power to have any sensitivity to new alleles.
We thus skip positions with fewer than 150 total observed sequences, summed over bins.
Positions with fewer than 30 observed mutations, also summed over bins, are similarly skipped.

For each \nsnp, we also ignore positions that do not have at least eight observed mutations in the \nsnp\nth\ bin.
This bin is of primary importance, because it is the means by which we determine that this \nsnp\ is the correct one, rather than those slightly larger or smaller.
If this bin is truly signaling a new allele, then it must contain a significant number of mutated sequences.

For several of the subsequent steps, we use an approximate fitting procedure to arrive at a slope, intercept, and associated uncertainties.
While less accurate, and more heuristic, than the least-squares fits that are used elsewhere, it is also much faster.
We begin by calculating the two-point slope between each pair of adjacent points.
If there are only two points in total, this is supplemented by a ``synthetic'' slope between the first point shifted up, and the second shifted down, by their respective uncertainties.
The approximate slope is then calculated as the mean of these pairwise slopes, with its uncertainty the associated standard error.
We arrive at the approximate y-intercept with a similar procedure, except that the pairwise slope is replaced by the pairwise y-intercept, which uses the previously-calculated pairwise slope.

For smaller \nsnp\ (three or less), we also require that the approximate upper-region y-intercept fit bounds do not include zero.
If they do include zero, there will not be a significant difference between the one- and two-piece fits.
As an additional, and more stringent, test that the upper-region y-intercept for these smaller \nsnp\ is well above zero, we require that the approximate fit's y-intercept is also greater than zero.

For \nsnp\ equals two, we also require that the bin immediately before the \nsnp\nth\ bin be outside of the upper-region y-intercept fit bounds.

And finally, for larger \nsnp\ (five or greater), the approximate lower fit's slope must be less than that of the approximate upper fit, in cases in which they are inconsistent.

\subsection*{Initial germline database}
To construct the initial germline database, we begin by taking the functional V, D, and J genes from the \imgt\ database.
We then also remove non-full length V sequences.

\subsection*{Excluded bases on 5' and 3' ends}
We are typically analyzing only partial V sequences, which leads to additional complications.
On the 5' end, the method must account for samples in which the read does not extend through the entire V gene.
On the 3' end, meanwhile, VDJ rearrangement itself deletes some number of bases.
The presence of incomplete V sequences clearly reduces our sensitivity to new alleles simply by reducing the sample size for positions at each end.
A more serious problem, however, is that differing lengths cause sequences to be assigned to incorrect bins, since their apparent number of mutations is different than their true number.
In order to avoid this problem, we ensure that all analyzed sequences begin and end at the same aligned germline bases.
To accomplish this, for each end (5' and 3'), we find the deletion length such that only a small fraction ($f$, by default 0.01) of sequences have a longer deletion length.
We then exclude the fraction $f$ of sequences that have longer deletions.
Among the remaining sequences, we then exclude from the analysis the positions that fall within these deletion lengths.
For example, if 99\% of sequences have 3' deletions of four or fewer bases, then we would discard sequences with more than four 3'-deleted bases, and would not use that gene's four most-3' positions in the fitting procedure.
Note that on the 3' side of V, this exclusion procedure is especially important because the final few germline-aligned positions next to any non-templated insertion always have very poorly-measured mutation frequencies.

\subsection*{Collapsing clones}
Our method requires that we consider only independent mutation events, excluding any mutations that share a common ancestry.
In order to satisfy this requirement we attempt to select from each clone the largest possible set of sequences without shared mutations.
In doing this, we give preference to relatively unmutated sequences, since most new alleles are separated by only a few SNPs from known alleles.
Specifically, we sort the sequences from each clonal family in order of increasing apparent V mutation.
We then traverse this list, selecting each sequence that does not share any mutations with a previously-selected sequence.
As shown in~\cite{partis-clustering} it would be straightforward to use the full partis method to separate the sequences into clonal families.
However, for the purposes of ensuring independent mutations, there is little benefit to having precisely accurate clusters, since a slightly inaccurate clustering only results in slightly inaccurate uncertainties in the fits, and uncertainties on uncertainties are in practice never large enough to impact the analysis.
For the sake of speed, then, we simply cluster using inferred naive sequences, i.e.\ every sequence that is inferred to have the same naive sequence is clustered together.
This has the additional benefit of becoming more conservative as the sample size becomes large -- in other words it tends to over-cluster more as the space of potential naive rearrangements fills up, and nearby rearrangement events have very similar naive sequences.
This has the effect of sacrificing some sensitivity in order to ensure that mutations are actually independent.

\subsection*{Initial removal of less-likely alleles}
Some care is necessary when constructing each sample's initial set of known genes.
We find the performance of our new-allele inference to be robust enough that the best approach is to first choose a minimal number of genes whose presence in the sample is supported by very strong evidence.
We then apply the new-allele inference framework in order to reinstate alleles for which the evidence was less overwhelming, along with any novel alleles.

In order to construct each sample's minimal initial gene set, we first partition the complete set of \imgt~\cite{Lefranc2009-iu} genes into groups within which SHM can easily cause confusion, and then retain only the most common gene in each group.
Note that this partitioning cannot be accomplished using only the \imgt\ names -- there are many cases of allelic variants that differ by so many SNPs that confusion is very unlikely, as well as alleles of separate genes that differ by only a single SNP.
We construct these groups by single-linkage clustering such that genes with the same conserved cysteine position, and separated by fewer than eight SNPs, are grouped together.
In order to ensure that we can re-infer all of the genes within each group, this number corresponds to the maximum number of SNPs for new-allele inference.

We also discard alleles that appear to occur at extremely low frequencies, by default less than one part in 2000.


\subsection*{Template allele removal}
The procedure outlined thus far can yield a confident judgment on whether there exists a previously-unknown allele separated from some known, ``template'' allele.
We must also, however, distinguish between cases where this template allele is also present in the sample, and cases where it is not (and was simply the closest known allele).
In order to do this we observe that in the plots in Fig~\ref{FIGsimpleFits} the y-intercept of the upper (post-\nsnp) fit is determined largely by the prevalence of the new allele.
For \nsnp\ near one, the y-value is very close to the actual allele prevalence, while for larger \nsnp\ the relationship is more complex.
When the new allele's prevalence is 1, i.e.\ the template allele is not present in the sample, however, the fitted post-\nsnp\ y-value is also very close to 1.
The only deviation is a slight decreasing slope from reversion to germline at higher mutation levels.
We thus remove template genes from the germline set when the upper fits for each position have y-intercept $1.1\pm 0.12$ and slope $-0.01\pm 0.015$.

\subsection*{Adding a new allele}
Once we have decided that there is sufficient evidence for a new allele separated by \nsnp\ SNPs from an existing allele, there remain several additional considerations.

First, we must determine its original germline sequence.
We begin by restricting ourselves to sequences assigned to the \nsnp\nth\ bin, i.e.\ which contain \nsnp\ apparent mutations.
This restriction is important, because unmutated sequences stemming from the new allele are assigned to this bin.
It thus minimizes the confusion caused by mutated sequences derived from the existing allele, as well as from any additional new alleles.
For each of the \nsnp\ positions where this allele differs from the template allele, we then choose as the new allele's germline nucleotide the most commonly-observed non-template nucleotide at that position.

If the newly-identified allele was present in the original, full germline set, then we add it with its original name; otherwise we add it with a provisional name derived from the template gene.
Because of the unavoidable ambiguity created by 3' exonuclease deletion (and short reads), in order to be considered equivalent we require only that two alleles are identical after applying the 5' and 3' exclusions described above.
If, for instance, we infer a new allele that differs by several SNPs from some template gene, and there is an existing allele in the original set that is identical to this new allele except for an extra base to the 3' of the cysteine, we assume that the newly-inferred and existing alleles are in fact the same.
More generally, we note that any new-allele inference framework that uses expressed data will suffer from a large uncertainty as to the precise number and identity of a V gene's most-3' few germline bases.
In order to resolve this uncertainty we must perform germline sequencing.

\subsection*{Simulation details}
The simulated samples used for validation were made with the same basic framework described in~\cite{partis-annotation}.
In addition to the details described there, we have added options to control various aspects of a sample's germline set.
All simulation options are described in detail in the manual (\url{https://git.io/vFKok}).

Most basically, we have added the ability to insert into a germline set new alleles that are separated from existing alleles by both point and insertion/deletion mutations.
The number of each type of mutation, and their properties, are specified with command line options.
Each mutation occurs either at a specified position in the allele's sequence, or at a random position within specified bounds (for instance, within vs outside of the CDR3).
These options allowed the creation of the simplified sparse gene repertoire samples in the Results.

These sparse repertoires are built around a single known germline gene.
We then add either one or two novel alleles, separated by SNPs at uniformly-selected random positions, from this existing germline gene.

In order to generate a germline set for the full repertoire samples, for each region we first choose some number of genes from the \imgt\ set, and then some number of alleles for each of these genes.
The mean number of alleles per gene is specified on the command line, then for each gene we choose a number of alleles from $\{1, 2\}$ with weights such as to (on average) arrive at the specified mean over all genes.
For both the ``low-SHM'' and ``high-SHM'' full-repertoire samples, this procedure was followed with 42, 22, and 6 genes (V, D, and J regions), with a mean alleles per gene of 1.33, 1.1, and 1 (V, D, and J).
This is concordant with the references in Results above, in particular~\cite{boyd2010-yj}, which reported a mean over 12 individuals of 40.2 homozygous, 8.6 two-allele heterozygous, and 1.1 three-allele heterozygous V genes, for an overall mean alleles per gene of 1.2.
Six novel alleles were then added, separated by 1, 1, 2, 3, 5, and 6 point mutations (at uniformly-selected random positions) from an existing allele.

We choose each gene's relative prevalence counts from a uniform random distribution with bounds $[1, 1/f_{\min}]$, where $f_{\min}$ is the minimum desired prevalence ratio between any pair of genes in the repertoire.
This ensures that the prevalence ratio for every pair of genes in the repertoire is in $[f_{\min}, 1]$, where $f_{\min}$ equals 0.15 (``low-SHM'' samples) or 0.05 (``high-SHM'' samples).
While this is roughly compatible with the variation in expression levels typically reported in real data, we emphasize that most previous studies (including our own~\cite{partis-annotation}) have aligned against the full \imgt\ set, and as such their reported expression levels for less-common genes are probably meaningless (Table~\ref{TABLEglsSimSummary}, Fig~\ref{FIGglsSimFull}).

\begin{figure}[!ht]
\forarxiv{\includegraphics[width=5.5in]{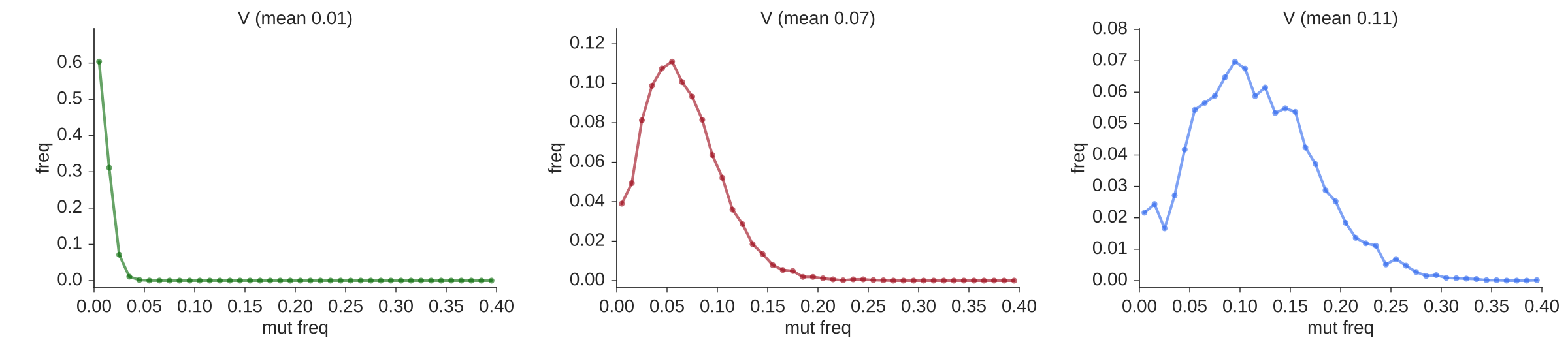}}
\caption{\
  {\bf V region mutation frequency distributions} from the low (left), typical (middle), and high (right) mutation samples used to make the top panels of Figure~\ref{FIGvarval1}.
  In the context of these distributions, the full-repertoire samples (Figs\simGlsTrees) correspond to a mean value of 0.02 (``low-SHM'' samples) and 0.06 (``high-SHM'' samples).
}
\label{FIGvFreqDistr}
\end{figure}

The SHM distributions in the full-repertoire samples were chosen to be representative of typical IgM-specific data (``low-SHM'', mean value 0.02) or typical unsorted samples (``high-SHM'', mean value 0.06) (compare to mean values in Fig~\ref{FIGvFreqDistr}).

Finally, we must decide on the clonal family structure of each sample.
Real repertoires vary widely in both their clonality and lineage structure.
However, we have shown in \varvalfigs that our clonal family collapse is an effective-enough approximation that changes in clonality and lineage structure are equivalent to changes in sample size, and thus only affect sensitivity.
In order to maximize the variety of interesting variables over which we can perform validation, we thus simulate the full repertoire samples with singleton clonal families.

\subsection*{Phylogenetic comparison plots}
In order to make the phylogenetic gene set comparison plots (Figs\simGlsTrees\methodVsMethodFigures\sampleVsSampleFigures) we begin by aligning all the genes that we want to compare using MUSCLE~\cite{Edgar2004-wt} (v3.8.31 with default parameters).
We then use RAxML~\cite{Stamatakis2014-lg} (v8.2.10, with the GTR model) to create a tree for these genes.
In order to allow easy visual comparison of the entire germline gene set in one plot, while also allowing comparison within each gene family (\imgt\ definition, e.g.\ IGHV3), we then collapse to length zero each branch that joins two different gene families.
If the reader would like to compare combinations of germline sets that are not shown in this paper, all true and inferred germline sets, for simulation and data, are available at \zenodolink, and the command line script used to make these plots is available at \url{https://git.io/vFo2B}.

\section*{Acknowledgements}

The authors would like to thank Kristian Davidsen, Laura Noges, and Peter Ralph for helpful discussions.
\forarxiv{
This research was supported by NIH grants U19-AI117891, R01-GM113246, R01-AI120961, and R01-AI138709.
The research of Frederick Matsen was supported in part by a Faculty Scholar grant from the Howard Hughes Medical Institute and the Simons Foundation.
}

\bibliographystyle{plos2015}
\bibliography{germline-set-generation}

\clearpage

\section*{Supplementary Information}


\subsection*{Comparison between methods}

\beginsupplement

\begin{table}[!ht]
  {\small
    \centering
    \begin{tabular}{cccccc}
   authors              &  year  &  citation                      &  ENA/SRA number  &  ethnicity  &  sequencing methodology  \\
\hline
   Vander Heiden et al  &  2017  & \cite{Vander_Heiden2017-fs}    &   PRJNA338795    &     n/a     &  MiSeq                   \\
   Laserson U et al     &  2014  & \cite{Laserson2014-yh}         &      n/a         &     n/a     &  454                     \\
   Sheng Z et al        &  2017  & \cite{Sheng_Schramm_Kong_NISC} &   SRP067168      &     n/a     &  MiSeq/454/5'RACE        \\

\hline
\end{tabular}
\caption{\
  {\bf Summary of data samples used.}
}\label{TABLEdataSamples}}
\end{table}

While a comprehensive comparison of the details of all three methods (\tigger, \igdiscover, and partis) is beyond the scope of this paper, we highlight some instructive details.
First, the default partis method and \tigger\ are more similar to each other than either is to \igdiscover.
Both use a fitting procedure on mutation accumulation plots to look for new alleles.
However, as described above, our approach to extracting information from these plots is very different, using hypothesis comparison rather than sharp cutoffs, for instance on the y-intercept.

\igdiscover, on the other hand, takes a quite different approach, clustering together sequences by distance and taking the consensus sequence of each cluster to be a germline gene.
The main advantage of this approach is that it enables detection of new genes which are separated by either point mutations or insertions/deletions from existing alleles.
The purely mutation accumulation plot-based approaches employed by \tigger\ and default partis, in contrast, can only detect new alleles separated by point mutations.
The tradeoff is that the fitting-based methods are able to use more detailed position-based information which allows them to function well in repertoires with higher SHM.
As noted by the \igdiscover\ authors, distance-based clustering methods, in general, suffer from the fact that once SHM is high enough that clusters from distinct V genes start bleed together, the heuristic thresholds used to separate clusters create significant inaccuracies.

On human repertoires, since the \imgt\ set is already fairly complete, the ability to detect new alleles separated by insertion/deletion mutations is not particularly important.
The germlines of most other species, however, are much less well characterized.
It is thus quite common in these species to encounter novel alleles that are not simply allelic variants of well-known genes.

The obvious course of action, then, is to combine the mutation accumulation plot-based and clustering-based methods in order to allow accurate inference on non-naive repertoires of all species.
We have, in fact, implemented this as a non-default option in partis (see Manual, at \url{https://git.io/vF12k}), which we recommend for non-human samples.
However, after extensive validation of this combined method, we believe that a somewhat modified clustering approach will be required to achieve better performance on highly mutated samples from all species, and thus leave its description to a future paper.


\newcommand{\noleafnameblurb}[1]{A version of Fig~\ref{#1} where leaves have been labeled with gene names.}

\begin{figure}[!ht]
\forarxiv{\includegraphics[width=5in]{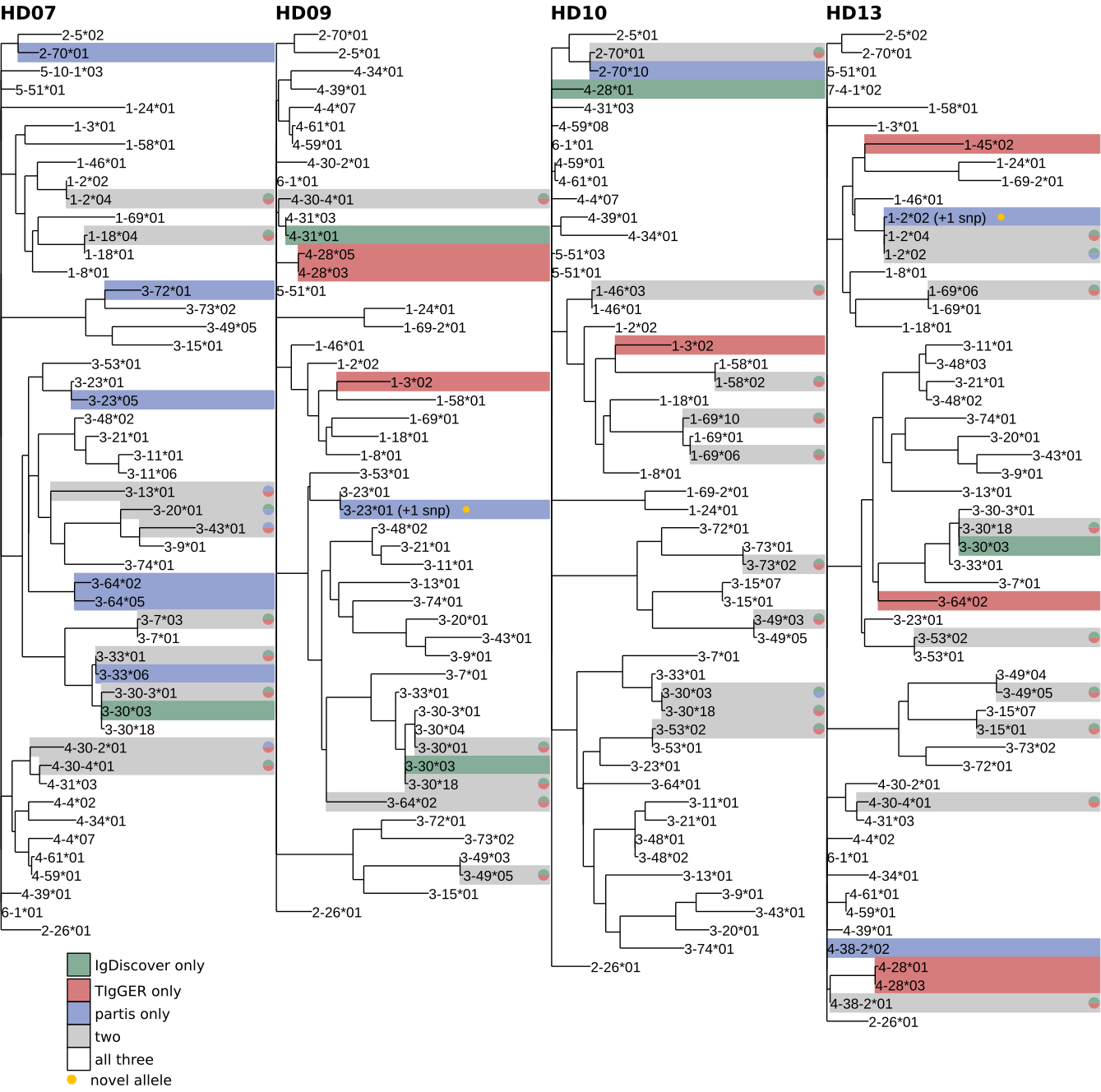}}
\caption{\
  \noleafnameblurb{FIGdataJasonMGHDmethodVsMethod}
}
\label{FIGdataJasonMGHDmethodVsMethodLeafNames}
\end{figure}

\begin{figure}[!ht]
\forarxiv{\includegraphics[width=3in]{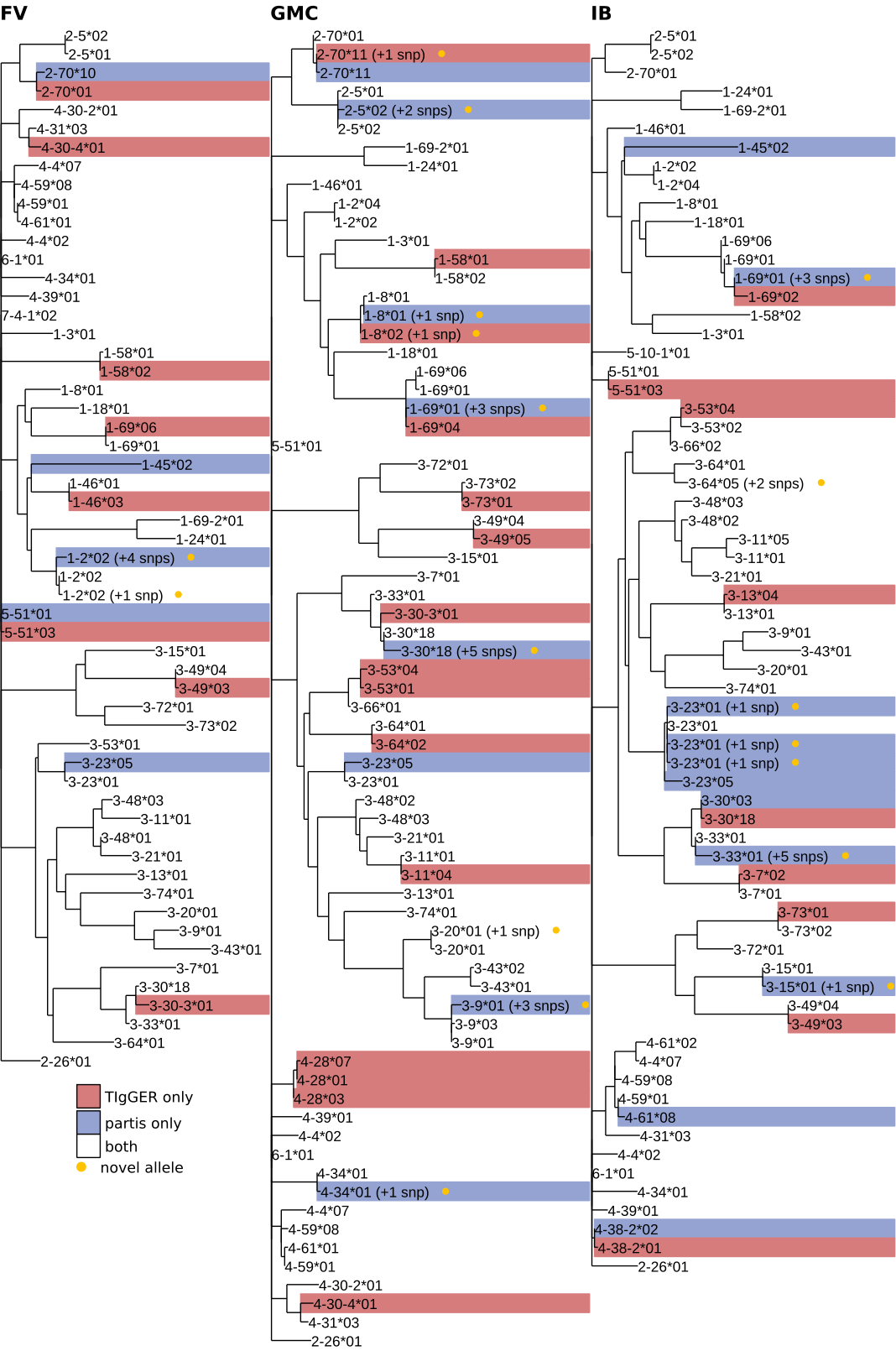}}
\caption{\
  \noleafnameblurb{FIGdataJasonInfluenzaMethodVsMethod}
}
\label{FIGdataJasonInfluenzaMethodVsMethodLeafNames}
\end{figure}

\begin{figure}[!ht]
\forarxiv{\includegraphics[width=4in]{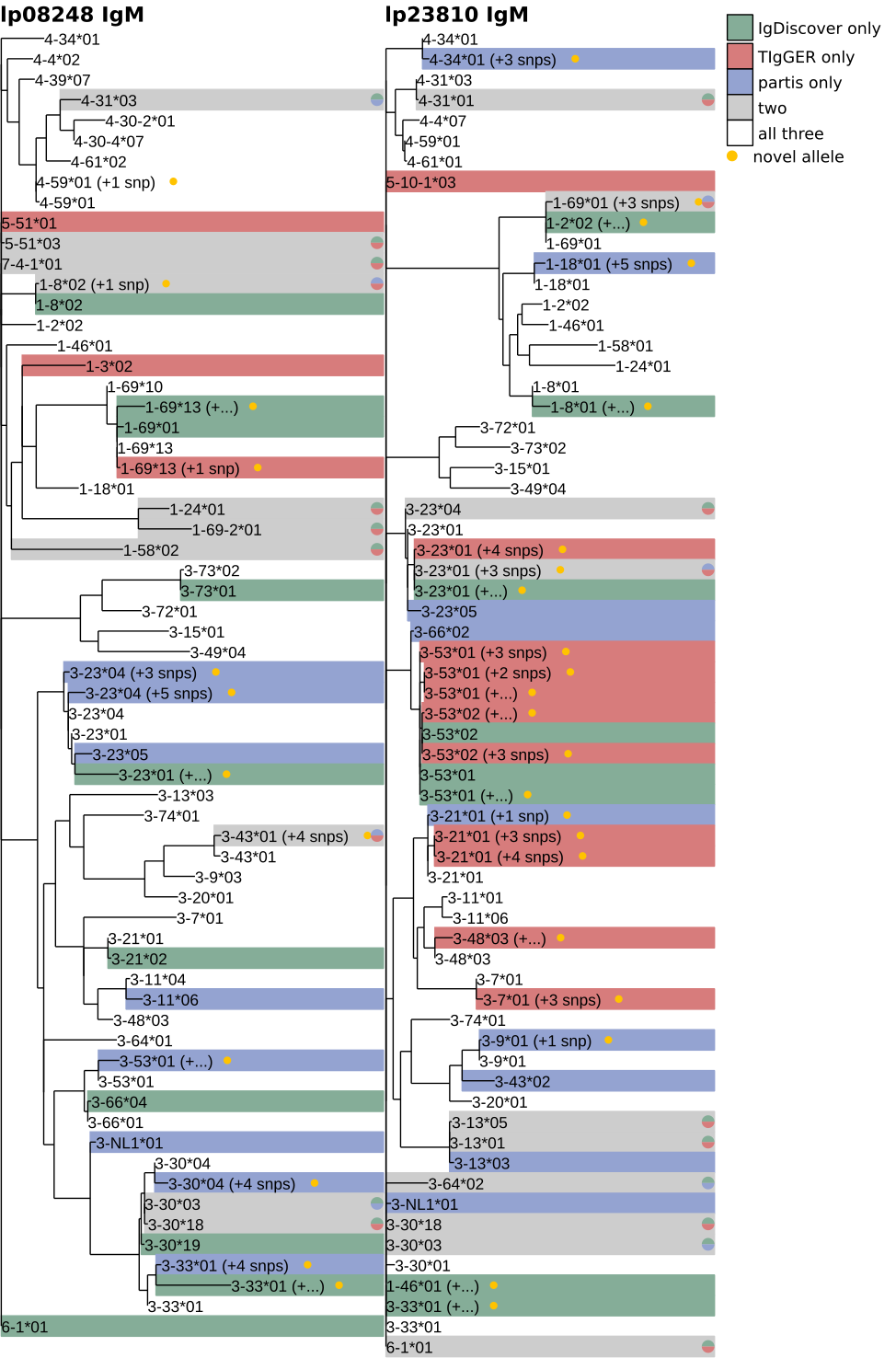}}
\caption{\
  \noleafnameblurb{FIGdataShengGSSPMethodVsMethod}
}
\label{FIGdataShengGSSPMethodVsMethodLeafNames}
\end{figure}

\begin{figure}[!ht]
\forarxiv{\includegraphics[width=4in]{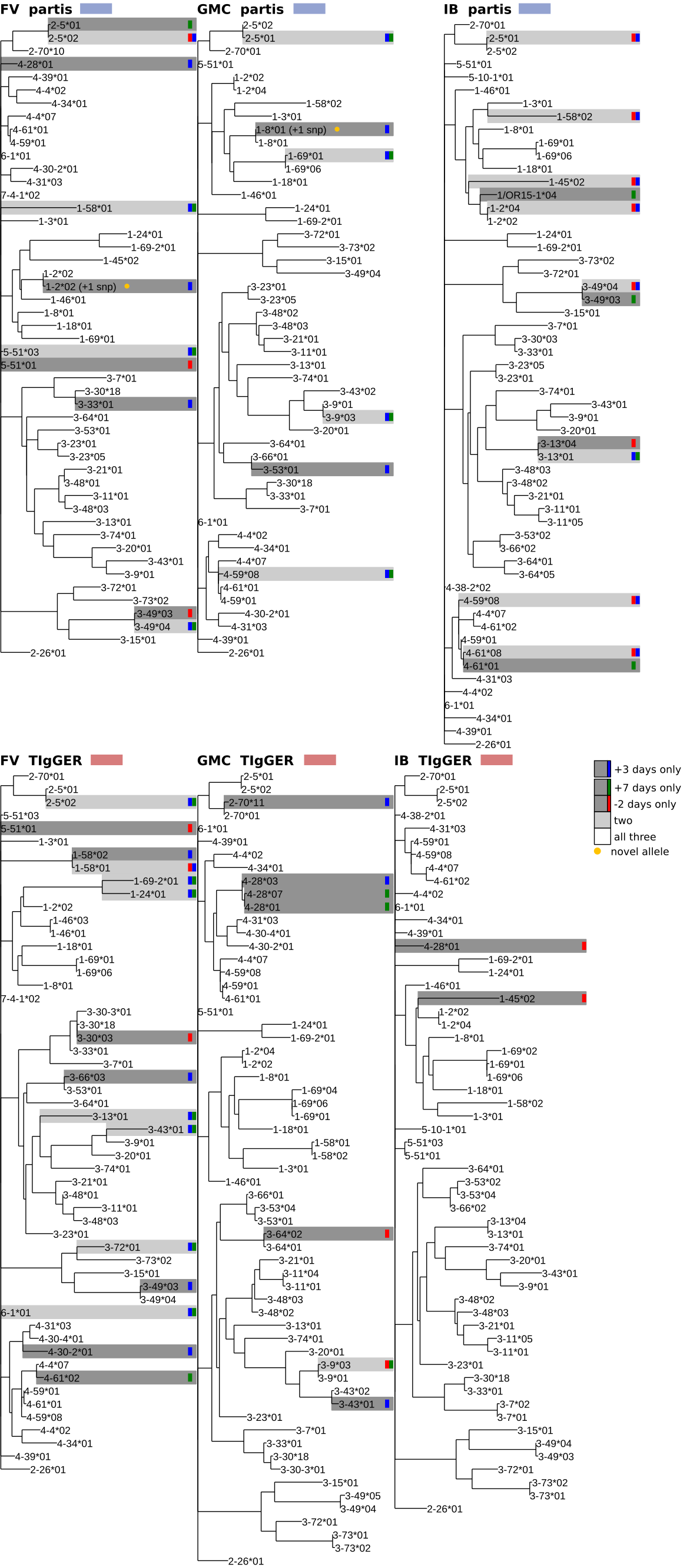}}
\caption{\
  \noleafnameblurb{FIGdataJasonInfluenzaSampleVsSample}
}
\label{FIGdataJasonInfluenzaSampleVsSampleLeafNames}
\end{figure}

\begin{figure}[!ht]
\forarxiv{\includegraphics[width=5in]{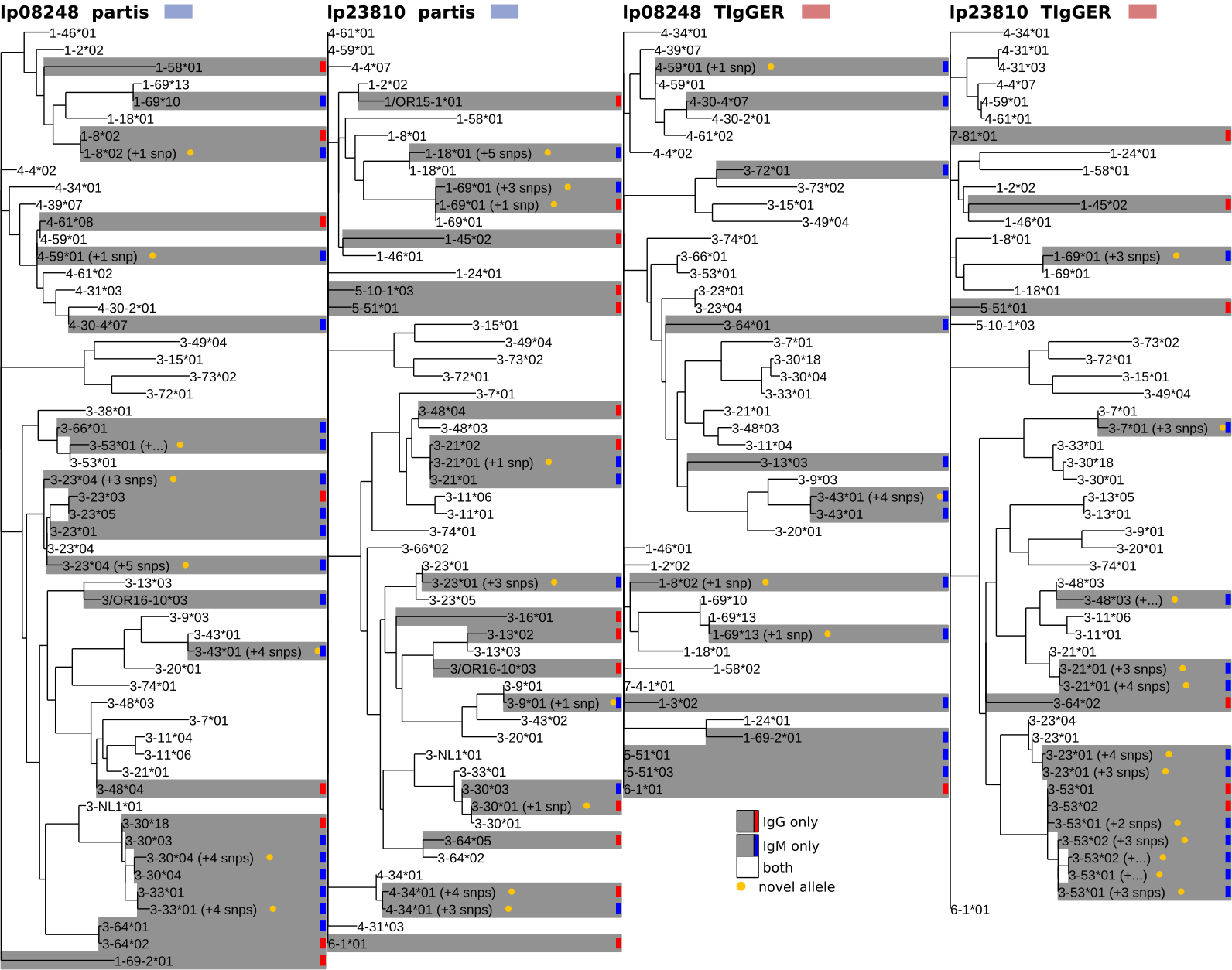}}
\caption{\
  \noleafnameblurb{FIGdataShengGSSPsampleVsSample}
}
\label{FIGdataShengGSSPsampleVsSampleLeafNames}
\end{figure}

\begin{figure}[!ht]
\forarxiv{\includegraphics[width=5in]{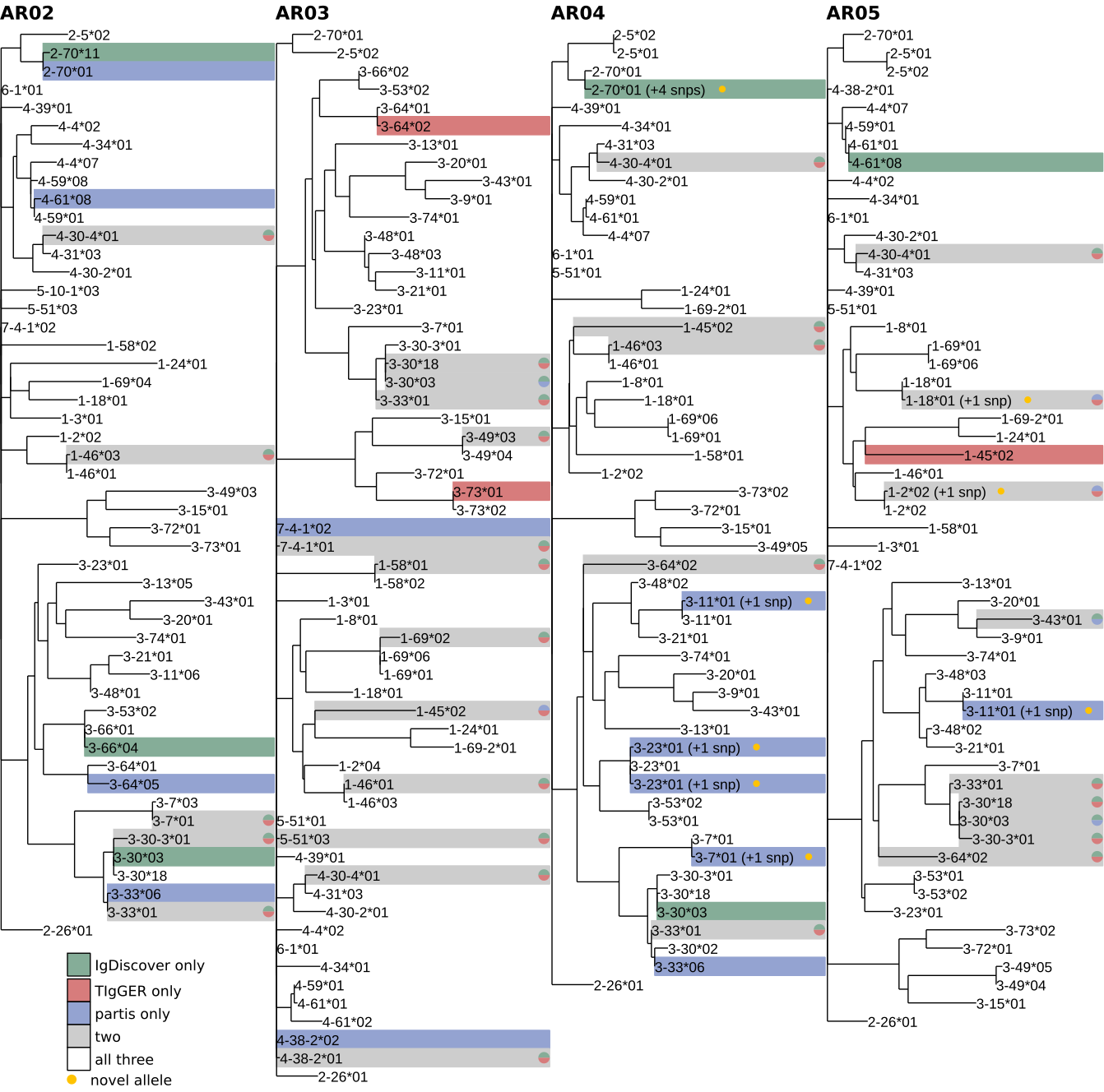}}
\caption{\
  \noleafnameblurb{FIGdataJasonMGARmethodVsMethod}
}
\label{FIGdataJasonMGARmethodVsMethodLeafNames}
\end{figure}

\begin{figure}[!ht]
\forarxiv{\includegraphics[width=5in]{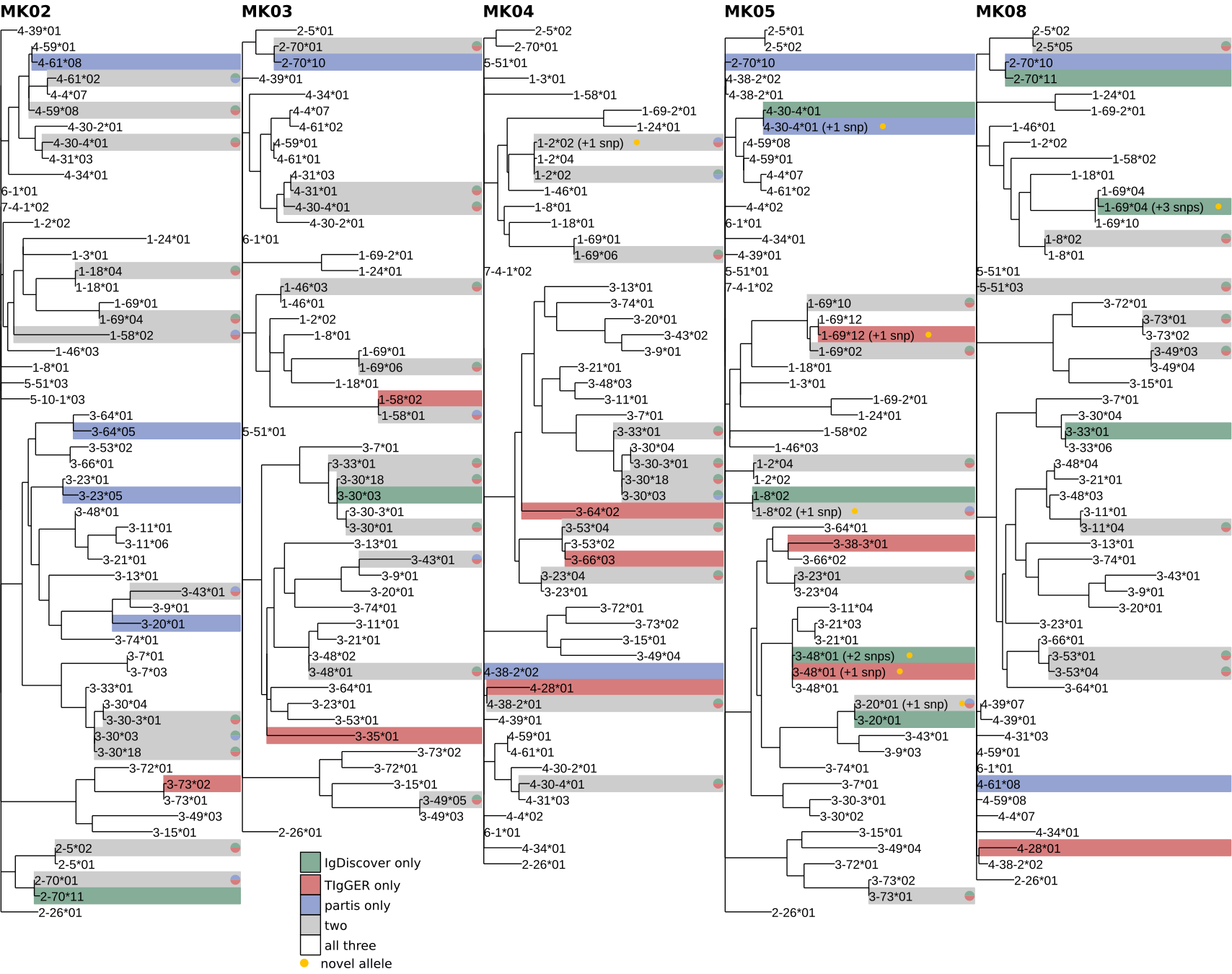}}
\caption{\
  \noleafnameblurb{FIGdataJasonMGMKmethodVsMethod}
}
\label{FIGdataJasonMGMKmethodVsMethodLeafNames}
\end{figure}

\end{document}